\documentclass{aa}  

\usepackage{txfonts,epsfig,graphicx,natbib,url,twoopt}
\usepackage{longtable}
\usepackage[breaklinks=true]{hyperref} %% to avoid \citeads line fills
\usepackage[labelfont=footnotesize,textfont=footnotesize]{caption}  %% as A&A

\bibpunct{(}{)}{;}{a}{}{,}    %% natbib format choice of A&A and ApJ
\newcommandtwoopt{\citeads}[3][][]{\href{http://adsabs.harvard.edu/abs/#3}%
                                        {\citealp[#1][#2]{#3}}}
\newcommandtwoopt{\citepads}[3][][]{\href{http://adsabs.harvard.edu/abs/#3}%
                                         {\citep[#1][#2]{#3}}}
\newcommandtwoopt{\citetads}[3][][]{\href{http://adsabs.harvard.edu/abs/#3}%
                                         {\citet[#1][#2]{#3}}}
\newcommandtwoopt{\citeyearrads}%
  [3][][]{\href{http://adsabs.harvard.edu/abs/#3}%
                                         {\citeyear[#1][#2]{#3}}}
\makeatother        

\begin{document} 
   \title{ A catalogue of Large Magellanic Cloud star clusters observed in the Washington photometric system}

   \author{T. Palma\inst{1,2,3} 
      \and 
      L.V. Gramajo\inst{3} 
      \and 
      J.J. Clari\'a\inst{3,4} 
      \and 
      M. Lares\inst{3,4,5} 
      \and 
      D. Geisler\inst{6} 
      \and 
      A.V. Ahumada\inst{3,4} }

\institute{Millennium Institute of Astrophysics, Nuncio Monse\~nor Sotero Sanz 100, Providencia, Santiago, Chile
\and
Instituto de Astrof\'isica, Pontificia Universidad Cat\'olica de Chile, Av. Vicu\~na Mackenna 4860, 782-0436 Macul, 
Santiago, Chile\\ \email{tpalma@astro.puc.cl} %\label{inst1}
\and
Observatorio Astron\'omico, Universidad Nacional de C\'ordoba, Laprida 854, 5000 C\'ordoba, Argentina 
\and 
Consejo Nacional de Investigaciones Cient\'ificas y T\'ecnicas (CONICET), Argentina
\and
Instituto de Astronom\'ia Te\'orica y Experimental (IATE), Laprida 922, C\'ordoba, Argentina
\and
Departamento de Astronom\'ia, Universidad de Concepci\'on, Casilla 160-C, Concepci\'on, 
Chile %\label{inst2}
}

\date{Received ; accepted }

% \abstract{}{}{}{}{} 
% 5 {} token are mandatory
 
  \abstract%
% context heading (optional)
   {} %leave it empty if necessary  
  % aims heading (mandatory)
   {The main goal of this study is to compile a catalogue including the fundamental parameters of a complete
    sample of 277 star clusters (SCs) of the Large Magellanic Cloud (LMC) observed in the Washington
photometric system, including 82 clusters very recently studied by us.}
  % methods heading (mandatory)
    {All the clusters' parameters such as radii, deprojected distances, reddenings, ages and metallicities
    have been obtained by appyling essentially the same procedures which are briefly described here.
    We have used empirical cumulative distribution functions to examine age, metallicity and deprojected
    distance distributions for different cluster subsamples of the catalogue.}
  % results heading (mandatory)
    {Our new sample made up of 82 additional clusters recently studied by us represents about a 40\% increase in the
    total number of LMC  SCs observed up to now in the Washington photometric system. In particular, we report here the
    fundamental parameters obtained for the first time for 42 of these clusters. We found that single LMC SCs are typically
    older than multiple SCs. Both single and multiple SCs exhibit asymmetrical distributions in log (age).
    We compared cluster ages derived through isochrone fittings obtained using different
    models of the Padova group. Although $t_G$ and $t_B$ ages obtained using isochrones from Girardi et al.
    (2002) and Bressan et al. (2012), respectively, are consistent in general terms, we found that
    $t_B$  values are not only typically larger than $t_G$ ages but also that Bressan et al.'s age uncertainties
    are clearly smaller than the corresponding Girardi et al. values.}
  % conclusions heading (optional), leave it empty if necessary 
    {}

   \authorrunning{T. Palma et al.}
   \titlerunning{Catalogue of LMC star clusters observed in the Washington system.}
   \maketitle

\section{Introduction}

The Magellanic Clouds have long been considered an ideal laboratory to study a variety of objects and phenomena 
in nearby galaxies. In particular, the Large Magellanic Cloud (LMC) has been a target of intensive research because of 
its proximity and its almost face-on position in the sky \citep{hz09}, which facilitates a detailed analysis of its 
stellar populations. Various studies of LMC star clusters (SCs) have shown that these stellar populations differ 
from Galactic SCs in their typical radii, masses, kinematics, age distribution, etc. The nature and cause of the so called cluster age-gap of approximately 3 to about 12 Gyr, with only a single cluster lying in this range in the LMC \citep{g97} still remains of great interest, even more so if it is taken into account that a variety 
of HST observations have revealed that the corresponding age-gap in the field stars does not exist \citep[e.g.,]
[]{holtz,smecker}. Unfortunately, this vast cluster age-gap does not allow us to use SCs to trace the chemical enrichment
 and star-formation history of the LMC during such a long period of quiescence. However, the great abundance of clusters outside of this gap make them excellent tracers of these quantities otherwise.\\

The Washington photometric system, originally defined to study G and K late-type stars and old stellar populations \citep{canterna}  and later calibrated by \citet{g91}, has been widely applied to young, intermediate-age and old clusters in the Galaxy \citep[e.g.,][]{cla07,p09a} and in the Magellanic Clouds \citep[e.g.][]{g03}. This system has proved to be an excellent tool to determine a variety of astrophysical parameters such as distances, interstellar absorption, ages and particularly chemical abundances (metallicities) for SCs as well as for field stars located in the cluster surrounding regions. The advantages that this system offers to study Galactic and/or extragalactic SCs have been demonstrated by \citet{g91,g97} and  \citet{gs99}. In particular, the combination of the Washington system $C$ and $T_1$ filters is approximately three times more metallicity-sensitive than the corresponding $VI$ standard giant branch technique. This, combined with the  Washington system's broad and efficient passbands, makes it a very powerful tool for exploring stellar populations in both nearby and especially more distant galaxies. In 
addition,  \citet{g87} has shown that the system can be made even more efficient by substituting the
Kron-Cousins $R$ filter for the standard Washington $T_1$ filter. In fact, the $R$ filter has a very similar wavelength coverage but a significantly higher throughput as compared to the standard Washington $T_1$ filter. Therefore, $R$ instrumental magnitudes can be easily and accurately transformed to yield standard $T_1$ magnitudes. As shown by \citet{gs99}, the combined use of the $C$ and $R$ filters allows us to derive accurate metallicities based on their standard giant branch (SGB) technique. Using LMC SCs observed in the Washington photometric system, we have recently examined the chemical enrichment history of the LMC during the last 2-3 Gyr \citep{palma15}. \\

The fundamental parameters of a total of 195 LMC SCs have been determined using the Washington
system in a number of papers over the years. A short description of the selected sample criteria is summarized in Table \ref{t:descrip}. We have recently published the results obtained for 40 unstudied or poorly studied SCs \citep{palma13,palma15}. In the 
current study, we report the results obtained for another 42 LMC SCs, all of them observed in the Washington system. Thus, this combined sample of 82 clusters represents about a 40$\%$ increase in the total of LMC clusters observed and studied up to this moment in the Washington photometric system. Since all the clusters' parameters were obtained by applying essentially the same procedures, this group of objects represents a uniform and homogeneous cluster sample. Although the total sample still represents only a tiny fraction of the very populous LMC SC system, it is in fact one of the largest and most uniform LMC SC samples available and thus a catalogue uniting the combined information in a single dataset should be of substantial general astrophysical use.\\

\begin{table*}[!ht]
   \caption{ Description of selected LMC SC samples found in the literature}
  \label{t:descrip}
  \centering
  \begin{tabular}{lclc}
  \hline
  Authors & Selected SCs & Characteristics / selection criteria  & CTIO Telescope\\
  \hline
  \citet{g87} & 1  & Technique tests & 1.5m \\
  \citet{g97} & 25   & Search for old SC candidates & 0.9m \\
  \citet{b98} & 13  & SC properties in the outer LMC & 0.9m \\
  \citet{p02} & 2  & Cluster age gap and first metallicity determination & 0.9m \\
  \citet{g03} & 8  & Metallicity determinations for some SCs from \citet{g97}  & 0.9m \\
  \citet{p03a} & 11  & Blue SCs in the west region of the LMC bar & 0.9m \\
  \citet{p03b} & 6  & Systematic study & 0.9m \\
  \citet{p09b} & 5  & Systematic study; mostly unstudied clusters & 0.9m \\
  \citet{p11} & 36  & Bursting forming episode &  4m\\
  \citet{palma11} & 4  & Systematic study; unstudied clusters & 4m \\
  \citet{p12a} & 26  & Enlarging the sample of SCs in the 100-1000 Myr age range & 4m \\
  \citet{palma13} & 23  & Systematic study; mostly unstudied clusters; age/metallicity gradients & 4m \\
  \citet{p14}  & 90  & Search for genuine SCs and age determinations & 4m \\ 
  \citet{palma15} & 17  & Systematic study; mostly unstudied clusters; age-metallicity relation & 4m \\
   \citet{choud} & 45  & Search for genuine SCs and characterization & 4m\\
  \hline
\end{tabular}
\end{table*}

\section{Observations}
We have compiled a catalogue including a total of 277 LMC SCs studied in the Washington system. All the photometric observations of these SCs were carried out at Cerro Tololo Inter-American Observatory (CTIO, Chile), using the Wahington $C$ and $T_1$ filters \citep{canterna} and the Kron-Cousins $R$ filter. As mentioned in \citet{palma15}, our new sample of 82 SCs was observed with the CTIO ``V\'ictor Blanco'' 4\,m telescope in December 2000. The CTIO 0.9\,m telescope was used by \citet{g97,g03}, \citet{b98} and \citet{p02,p03b,p03a,p09b,p11b}, while the observations reported by \citet{g87}, \citet{p11,p12a,palma13,palma15} and \citet{choud} were also carried out with the CTIO 4\,m telescope. Only one LMC cluster (NGC\,2213) was observed with the CTIO 1.5\,m telescope. The seeing at CTIO was typically 1-1.5 arcsec during all the observing nights. \\

\begin{figure}
\centering
\includegraphics[width=\columnwidth]{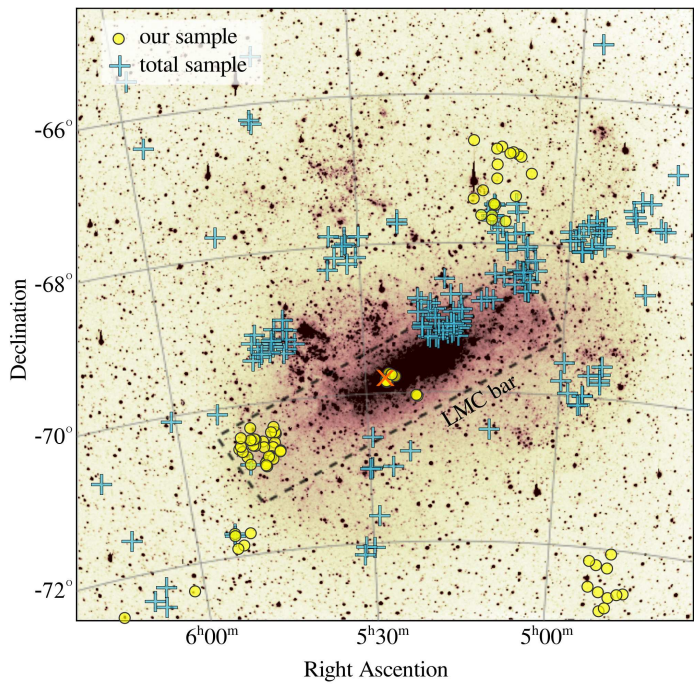}
\caption{ Distribution map of the whole sample of star clusters studied in the Washington photometric system 
superimposed on the LMC galaxy image. Filled circles represent the 82 clusters of our own sample, while plus signs stand for clusters studied by other authors using the same technique and analysis procedures. The cross indicates the geometrical centre of the bar \citep{bok}. The credits for the background image belongs to \citet{bothun}, \citet{kenn} and \citet{parker}.}
\label{f:map}
\end{figure}

The total cluster sample is presented in Table \ref{t:clusters}, where we list the various star cluster designations from 
different catalogues, 2000.0 equatorial coordinates, Galactic coordinates, and the cluster core radii given by \citet{b08}. 
These core radii constitute half of the mean apparent central diameters obtained by computing the average between the major (a) 
and minor (b) axes. Figure \ref{f:map} shows a distribution map of the whole sample of LMC clusters studied in the Washington photometric 
system superimposed on the LMC image. The dashed lines delimit the LMC bar region. Filled circles represent the 82 clusters 
of our recent sample, while plus signs stand for clusters studied by other authors using the same technique and analysis procedures. \\

\section{Fundamental parameters}

The procedures applied by both ourselves as well as other authors to determine the LMC SCs fundamental parameters in the Washington system are briefly described below. For further details see \citet{palma13} and references therein.\\

{\bf Cluster radii: } the procedure generally applied to determine a cluster's radius is based on obtaining the 
radial profile of the stellar density surrounding the cluster. Firstly, it is necessary to determine the cluster centre 
through Gaussian distribution fits to the star counts performed in the x and y directions.  The most important sources of uncertainty in the placement of cluster centres come from the relatively small ratio between the number of cluster and field stars, and the projected intracluster fluctuations due to both cluster and field star density variations. Then, the cluster radial profile is built by computing the number of stars per unit area at a given radius $r$.  Once the level of the background is estimated beyond the observed cluster boundary, the radius of the cluster is defined as the distance from the cluster's centre where the number of stars per unit area (cluster+background) equals that of the estimated background level. Error estimates for the radii range typically between 0.05' and 0.2', according to the telescope used, the concentration of the cluster and the position within the LMC. Examples of the procedure here described to determine clusters' radii can be seen in Palma et al. (2013).  In a few cases, when the clusters appear to be too faint if compared to the ``noisy'' background, the core radii reported by \citet{b08} were adopted.\\

{\bf Angular deprojected distances:} the angular deprojected distance of a LMC cluster is the distance, 
measured in degrees, from the optical centre of the LMC to the cluster, taking into account the 
depth of the LMC and is computed using the following expression \citep{cla05}:

\begin{equation}
\label{eq1}
d = d(p)\{1+[sin(p-p')^2][tg(i)^2]\}^{0.5},
\end{equation}

\noindent where $d$ is the cluster angular deprojected distance, $d(p)$ the angular projected distance on the plane of the sky, $p$ the position angle of the cluster (measured in the usual sense towards E starting from N), $p'$ the position angle of the line of nodes and $i$ the tilt of the LMC plane to the plane of the sky. The position of the cluster NGC\,1928 (J2000, $\alpha = 5^h 20^m 47^s\,\, \delta = -69^{\circ} 28' 41"$) was adopted as the LMC optical centre. To compute $d$ from equation (\ref{eq1}), we adopted  $i$ = 35.8$^{\circ} \pm$ 2.4$^{\circ}$ and $p'$ = 145$^{\circ} \pm$ 4$^{\circ}$  for the tilt of the LMC plane and the position angle of the line of nodes, respectively \citep{olsen}.  We carried out an error analysis in order to measure the uncertainties involved in the geometric parameters. We found that our estimated error in deriving this quantity increases with the angular projected distance reaching a maximum value of $\sim$ 0.3$^{\circ}$.\\

{\bf Interstellar reddening and cluster distances:} cluster-reddening values have usually been estimated by interpolating the extinction maps of \citet{bh82}. These maps were obtained from HI (21 cm) emission data for the southern sky and provided us with foreground $E(B-V)$ colour excesses, which depend on the Galactic coordinates. 
As shown in Col. 4 of Table \ref{t:clusters}, the resulting $E(B-V)$ values for the whole cluster sample range between 0.00 and 0.23, which are typical values for the LMC.  As explained in previous studies, we preferred not to use
the full-sky maps from 100-$\mu$ dust emission obtained by \citet{schlegel} since the dust temperature and reddening derived from pointing towards the LMC and other bright extragalactic sources are not reliable in these cases \citep[e.g.][]{p11b}. As for the cluster distance moduli, the value of the LMC true distance modulus $(m - M)_0$ = 18.50 $\pm$ 0.10 reported by \citet{s10} has always been adopted. According to \citet{subra}, the average depth for the LMC disc is 3.44 $\pm$ 0.16 kpc. Keeping in mind that any LMC cluster could be situated in front of or behind the main body of the LMC, we come to the conclusion that the difference in apparent distance modulus could be as large as $\Delta (V - M_V) \sim$ 0.3 mag. Given that the average uncertainty when adjusting the isochrones to the cluster colour-magnitude diagrams (CMDs) is 0.2-0.3 mag, adopting one single value for the distance modulus of all the clusters should not dominate the error budget in the final results.  \\

{\bf Ages and metallicities:} cluster ages and metallicities have been usually determined by applying two different and independent procedures. In both cases, however, it is necessary to first obtain the observed cluster $(C-T_1,T_1)$ CMD and then to minimize the field star contamination in this diagram. The CMDs of the 82 clusters of our recent sample were cleaned by using a statistical method developted by \citet{p12b}. Once the cleaned CMDs of the clusters are obtained, the first method consists in selecting a set of theoretical isochrones corresponding to different ages and metallicities and superimposing them on the cleaned cluster CMDs, once they were properly shifted by the corresponding $E(B-V)$ colour excess and LMC distance modulus. The age and metallicity adopted for each cluster are those corresponding to the isochrone which best matches the shape and position of the cluster main sequence (MS), particularly at the turn-off (TO) level, as well as the $T_1$ magnitude of the red giant clump (RGC). To apply this method, theoretical isochrones computed for the Washington system by the Padova group \citep{gir02,bressan} and Geneva group \citep{lesch01} have been used. \citet{lesch01} and \citet{gir02} isochrones have been computed for chemical compositions of $Z$ = 0.019, 0.008 and 0.004, equivalent to [Fe/H] = 0.0, -0.4 and -0.7, while \citet{bressan} more recent models include isochrones having metallicities between [Fe/H] = -0.19 and -0.84, which vary in an almost continuous way. Although for 23 out of the 82 clusters of our sample reported in \citet{palma13} we initially used \citet{gir02} isochrones, for the present analysis we have used \citet{bressan} isochrones for the entire sample because the latter include much smaller intervals in chemical composition ($Z$) than those used by \citet{gir02}. Thus, more precise fits could then be obtained. In general, the different sets of theoretical isochrones both from Padova and Geneva lead to nearly similar results. The differences arising when different sets of isochrones of the 
Padova group are used can be seen in Figure \ref{f:f5}.\\

A second method to derive cluster ages is based on the $\delta T_1$ parameter, defined as the difference in $T_1$ magnitude between the RGC and the MSTO in the Washington $(C-T_1,T_1)$ CMD. The age is obtained from the following equation given in \citet{g97}:

\begin{equation}
Age(Gyr) = 0.23+ 2.31\times \delta T_1 -1.80 \times \delta T_1^2 + 0.645\times \delta T_1^3,
\end{equation}

\noindent with a typical error of $\pm$0.3 Gyr. Age determination via $\delta T_1$, however, is applicable only to 
intermediate-age (IACs) and/or old clusters, i.e., generally older than 1 Gyr. Even though some clusters 
appeared to be IACs (1-3 Gyr), it was not possible to determine their ages from $\delta T_1$ due to the fact 
the RGC in their CMDs was not clearly visible. This happened because sometimes the central regions of the clusters seem to be saturated, or there are just very few RC stars in some faint clusters, or else they are not 
photometrically well resolved in the images. In these cases, the RG stars are missing or poorly defined and therefore 
no clump can be clearly detected in the CMDs. The resulting $\delta T_1$ values and the corresponding cluster 
ages are listed in Col. 7 of Table \ref{t:param}.  \\

Metallicities have also been obtained utilizing the SGBs of \citet{gs99} by placing the observations in the $[M_{T_1} , (C - T_1)_0]$ plane utilizing the equations $E(C-T_1)$ = 1.97$E(B-V)$ and $M_{T_1}$ = $T_1$ + 
0.58$E(B-V)$ - $(V-M_V)$. Each SGB corresponds to an iso-abundance curve. As these authors pose, however, this procedure can be applied only to SCs aged 2 Gyr or older. \citet{gs99} demonstrated that the metallicity sensitivity of the SGBs is three times higher than that of the V, I technique \citep{dacosta}. Consequently, it is feasible 
to determine metallicities three times more precisely for a given photometric error. The SGB method consists of 
inserting absolute $M_{T_1}$ magnitudes and intrinsic $(C-T_1)_0$ colours for the clusters into Fig. 4 of \citet{gs99} 
to roughly derive their metal abundances ([Fe/H]) by interpolation. The metallicities derived 
through this method were corrected for age effects for younger clusters, following the recommendations made by \citet{g03}. The resulting age corrected metallicities are shown in Col. 8 of Table \ref{t:param}, with typical errors of 0.3 dex. \\

\section{Catalogue description}

We have compiled a catalogue in which we included the fundamental parameters of 277 LMC SCs observed in the Washington photometric system. All these clusters have been studied and analyzed in a homogeneous way. The same procedures have been applied not only by our team but also by the other studies used in this compilation. The catalogue presented in Table \ref{t:param} is structured as follows: \\

\begin{itemize}
\item ID: cluster designations from different catalogues.
\item Cluster radius: distance $r$ in arcminutes from the cluster's centre up to the region where the stellar 
density equals that of the background.
\item Angular deprojected distance: distance, measured in degrees, from the LMC optical centre 
to the cluster.
\item E(B-V): cluster reddening.
\item Age$_I$: age in gigayears determined from isochrone fittings.
\item $[Fe/H]_I$: metallicity determined from isochrone fittings.
\item Age$_{II}$: age in gigayears derived from the $\delta T_1$ method.
\item $[Fe/H]_{II}$: metallicity estimated from the SGB procedure of \citet{gs99}.
\item Notes: letters a, b and c indicate that the clusters have also been studied in other photometric systems. 
Letter m denotes that the cluster is part of a binary or a multiple system \citep{dieball}. 
\item Ref: references to the works from which we took the fundamental parameters determined by other authors in the Washington photometric system.
\end{itemize}

\section{Statistical analysis}

\subsection{The empirical cumulative distribution function (ECDF)}

Although the age and metallicity distributions of the LMC SCs can be examined and compared by computing
their respective histograms, the corresponding empirical cumulative distribution functions (ECDFs) are preferred since they are independent of the particular selection of the binning function employed to build 
the histograms. The ECDF of a random sample of observations ${x_1, x_2, ..., x_n}$ is a function $F_n(t)$ 
given by the fraction of objects in the sample which are equal to or lower than an arbitrary value $t$. 
This function is equivalent to the original data. Its use avoids the loss of information brought about by the 
rounding off that results from placing the data in integer bin units when histograms are constructed \citep[e.g.][]{drion}.  \\

The ECDF can also be used to compare two samples and assess whether they are intrinsically different or just random realizations of the same parent distribution which appear different because of the stochastic nature of the data. The maximum difference between the two ECDFs, denoted by D, is commonly used as an indicator of the distinctiveness of the two data sets. This is the basis of the Kolmogorov-Smirnov test or KS-test \citep[e.g.][]{press}, which allows one to quantitatively contrast the hypothesis that the two samples are drawn from the same parent distribution to a given significance level, with the alternative hypothesis that the two samples are not taken from the same population. The statistic of this test has a known distribution that permits computing the p-value, i.e., the probability of obtaining, in two random samples, a value of the variable D equal to or larger than the one observed for the two particular data sets. Being nearly distribution free, this statistic is fairly easy to compute \citep{ross}.

\subsection{Statistical results}

We adopted the above defined ECDF to study the age and metallicity results obtained for different LMC cluster 
samples. The ECDF was also used to analyze the advantages of employing different sets of theoretical isochrones. 
Four different LMC SC samples are considered for our statistical analysis. The first of these samples,
called S0, includes all the SCs whose ages and metallicities have been determined by any other procedure. 
A second sample (S1) includes the 82 SCs of our own recent sample. A third (S2) is made up of all
277 SCs that have been observed and studied in the Washington photometric system up to this moment, i.e. this catalogue. 
A fourth sample (S3) represents the difference S2-S1. Table \ref{t:samples} shows the designations given to the 
different LMC cluster samples considered.  \\

\begin{table}[!ht]
\setcounter{table}{3}
   \caption{Designations of the different samples used}
  \label{t:samples}
  \centering
  \begin{tabular}{lcc}
  Name & Description & Number of SCs \\
  \hline
  S0 & Full LMC cluster sample  & 1970 \\
  S1 & Our cluster sample  & 82  \\
  S2 & Full Washington cluster sample & 277   \\
  S3 & S2-S1  & 195   \\
  \hline
\end{tabular}
\end{table}

\subsubsection{Metallicity distributions} 
\begin{figure} 
\centering
\includegraphics[width=\columnwidth]{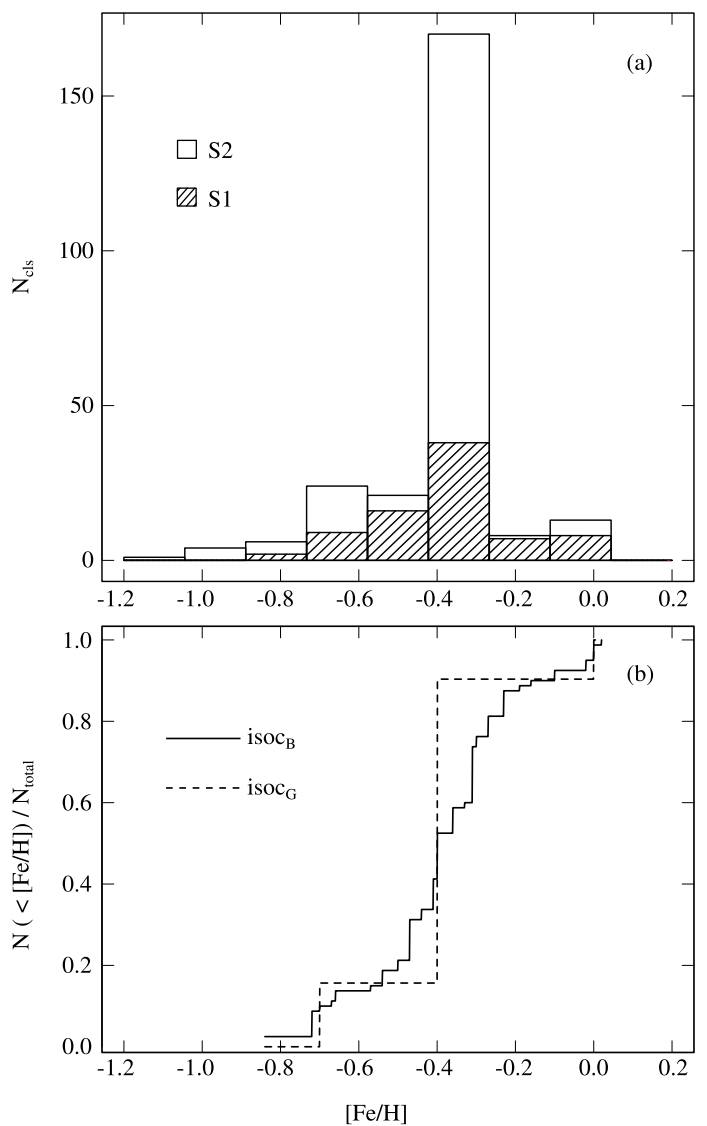}
\caption{Panel (a): histograms showing the metallicity distributions of LMC SCs in the S2 (empty bars) and S1 (dashed bars) cluster samples, respectively. Panel (b): comparison of metallicity ECDFs obtained for our own cluster sample S1 using sets of isochrones computed by \citet[dashed line]{gir02} and \citet[solid line]{bressan}.}
\label{f:f2}
\end{figure}

Panel (a) of Figure \ref{f:f2} exhibits the metallicity ([Fe/H]$_I$) histograms (number of clusters per metallicity interval) corresponding to cluster samples S2 (empty bars) and S1 (dashed bars), respectively. Although the metallicity distribution of the clusters belonging to S2 is narrower compared to that of cluster sample S1, the locations of the peaks of these two distributions are very similar within the errors ($<[Fe/H]> $= -0.39 and -0.42 for S2 and S1, respectively). In fact, a t-test for the difference of the mean metallicity values yields a result consistent with zero, to a statistical significance level of 95\%. Given the statistical evidence, it is not possible to conclude that there is a significant difference between samples  S1 and S2. This suggests that S1 and S2 come from the same parent distribution as far as metallicity is concerned. Also, the kurtosis is positive for the two samples (5.5 and 3.2 for S1 and S2, respectively), which indicates that although the S2 distribution is more concentrated around its mean value, both S1 and S2 are distributions narrower than what would be expected from a Gaussian distribution. More than 60\% of the clusters in S2, for example, have metallicities in the [-0.45, -0.35] range, in contrast with less than 30\% of the clusters in S1, for the same metallicity range. This difference may be due to the fact that for most of the clusters in the S2 sample, metallicities have a fixed value of -0.4 dex. In panel (b) of the same figure, the metallicity ECDFs obtained using \citet{gir02} and \citet{bressan} isochrone sets are shown by dashed and solid lines, respectively. Note that \citet{gir02} models only admit the adoption of discrete metallicity values (-0.7, -0.4 and 0.0 in our case). Note as well that there is a large jump in the metallicity ECDF at [Fe/H] = -0.4, as this is precisely the most frequent value. On the other hand, the metallicities obtained by using \citet{bressan} isochrones are distributed over different values in the same metallicity range. In this case, the metallicity ECDF is represented by a smoother curve and the metallicity mode is equal to -0.36. The difference between this value and the above mentioned -0.4 exceeds the mean metallicity error as a consequence of the procedures used to estimate the metallicities from isochrone fittings. \\

\subsubsection{Age distributions} 
\begin{figure}
\centering
\includegraphics[width=\columnwidth]{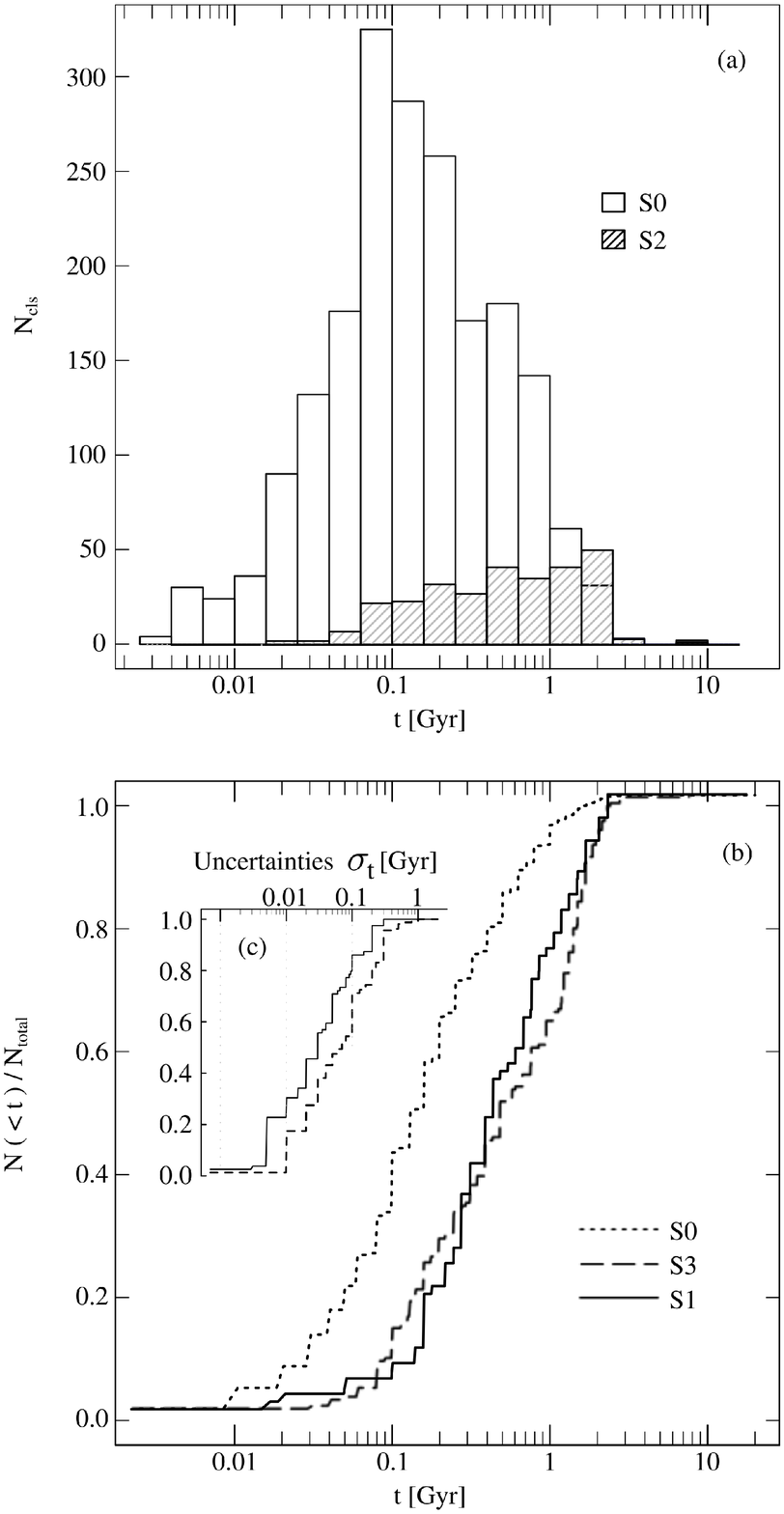}
\caption{Panel (a): histograms showing the age distributions obtained for the S0 (empty bars) and
S2 (dashed bars) cluster samples. Panel (b): age ECDFs for S1 (solid line), its complementary sample S3 
(dashed line), and S0 (short dashed line) cluster samples. The cumulative distribution estimates of age 
uncertainties are also shown inset in the figure.}
\label{f:f3}
\end{figure}

The age histograms and age ECDFs for different LMC cluster samples are shown in Figure \ref{f:f3}. In the upper panel (a), S0 and S2 samples are represented by empty bars and dashed bars, respectively. It can be clearly seen that the ages at which these two distributions reach their peak values are different. Indeed, the corresponding age of the S2 peak distribution is about 1.90 Gyr larger than that of S0. However, the shapes of these two distributions are fairly similar, since their skewness values are  nearly the same. Indeed, while 67\% of the clusters in S2 are younger than 1 Gyr, this percentage in the full cluster 
sample S0 rises to 90\%. This difference may result from the differing selection of targets made for each sample. In fact, most of the clusters included in S0, for example, have been studied by \citet{pu00} and \citet{glatt} who selected clusters younger than 1 Gyr. The age shift could also be partially reflecting the fact that cluster ages in S1 and S3 were estimated only from Washington photometric data, while those in S0 were obtained using different photometric systems (UBV, Washington, etc.). Age dispersions are also different, being 10\% larger in S0 than in S2. To quantify this difference, we performed a test for the ratio of variances of these two distributions, thus obtaining a difference of 1 to a statistical significance level of 95\%. In the lower panel (b) of Fig. \ref{f:f3}, different age ECDFs are 
shown. In the inset plot, the age error distributions for S1 and S3 samples are presented. Note that the age uncertainties are clearly smaller in the S1 sample. Indeed, 80\% of S1 clusters have uncertainties lower than 0.08 Gyr, while the lower tail of the uncertainties in the S3 sample reaches 0.3 Gyr. The mean values of these two distributions differ by 0.06 Gyr, which implies that the mean age uncertainty in S3 is more than twice the value of S1. A KS-test rejects the possibility that both distributions are consistent with a p-value of 0.003. \\

\subsubsection{Age and characteristics of single and multiple star clusters} 
\begin{figure}
   \centering
   \includegraphics[width=\columnwidth]{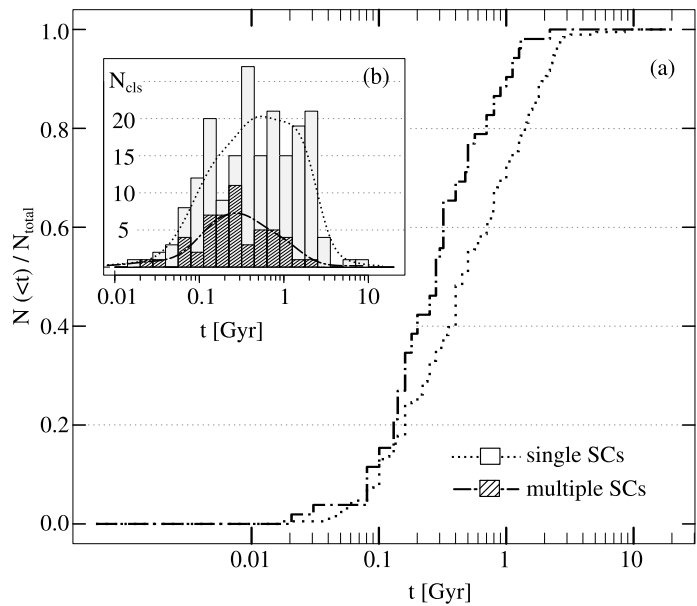}
   \caption{Age distributions for multiple and single LMC clusters. The empirical cumulative distributions for multiple systems (dashed dotted line) and single SCs (dotted line) are shown. On the inset panel (b),  the corresponding age histograms are shown (empty and shaded bars for single and multiple SCs, respectively) together with the kernel density estimations of both distributions (solid lines).}
   \label{f:f4}
   \end{figure}

We also examined the differences existing between the ages of single and multiple systems in the LMC. According to \citet{dieball}, multiple systems appear to be formed by close pairs or more clusters on the plane of the sky due to projection effects. In panel (a) of Figure \ref{f:f4}, the ECDFs for multiple (dashed dotted line) and single SCs (dotted line) are plotted. The corresponding age histograms as well as the kernel density estimations for these two distributions are also shown in the inset panel (b). It can be clearly seen in Fig. \ref{f:f4} that single LMC SCs are typically older than multiple SCs. This difference is reflected both in the means and the shapes of the distributions. The Welch t-test \citep{welch}, designed to test if two samples have the same mean, is more reliable than the Student t-test applied when the two samples have different sizes and unknown variances which are suspected of being also different. We used the Student-test to measure the difference in the age values of single and multiple systems. We found that the statistic for the difference of the sample means belongs to the confidence interval (0.11, 0.37) to a statistical significance level of 95\%.  The mean age values are 0.48 and 0.27 Gyr for single and multiple star cluster samples, respectively.  The F-test \citep{snede} is 
useful to determine if two populations have similar dispersions. The F-test uses the ratio of the sample variances as the test statistic, which follows an F distribution and thus permits testing the hypothesis that the ratio of the variances equals one. Although the variance for the sample of single SCs is approximately 40\% larger than that of multiple SCs, the F statistic for the quotient of the variances is not conclusive when discarding the null 
hypothesis of the variances being equal ($v_1/v_2$ = 1.43, p = 0.132). Besides, both multiple and single SCs exhibit asymmetrical distributions in $log(age)$, the latter being slightly more asymmetric than that of the multiple SCs. This is quantified by the skewness values of -0.212 and -0.196 obtained for single and multiple SCs, respectively, meaning 
that the tail of older clusters appears to be flatter and falls more abruptly. In spite of the distribution shapes appearing wide, kurtosis values are 2.42 and 3.02 for simple and multiple SCs, respectively. These values indicate that the tails of the distributions are truncated with respect to a Gaussian distribution. The kernel density estimations of the two age distributions are also shown  in panel (b) of Fig. \ref{f:f4}. This does not depend on the bin size used to build the age histogram and contributes to enhance the distribution asymmetry, especially for single SCs. \\

\subsubsection{Age determinations with different models} 
\begin{figure} 
   \centering
   \includegraphics[width=\columnwidth]{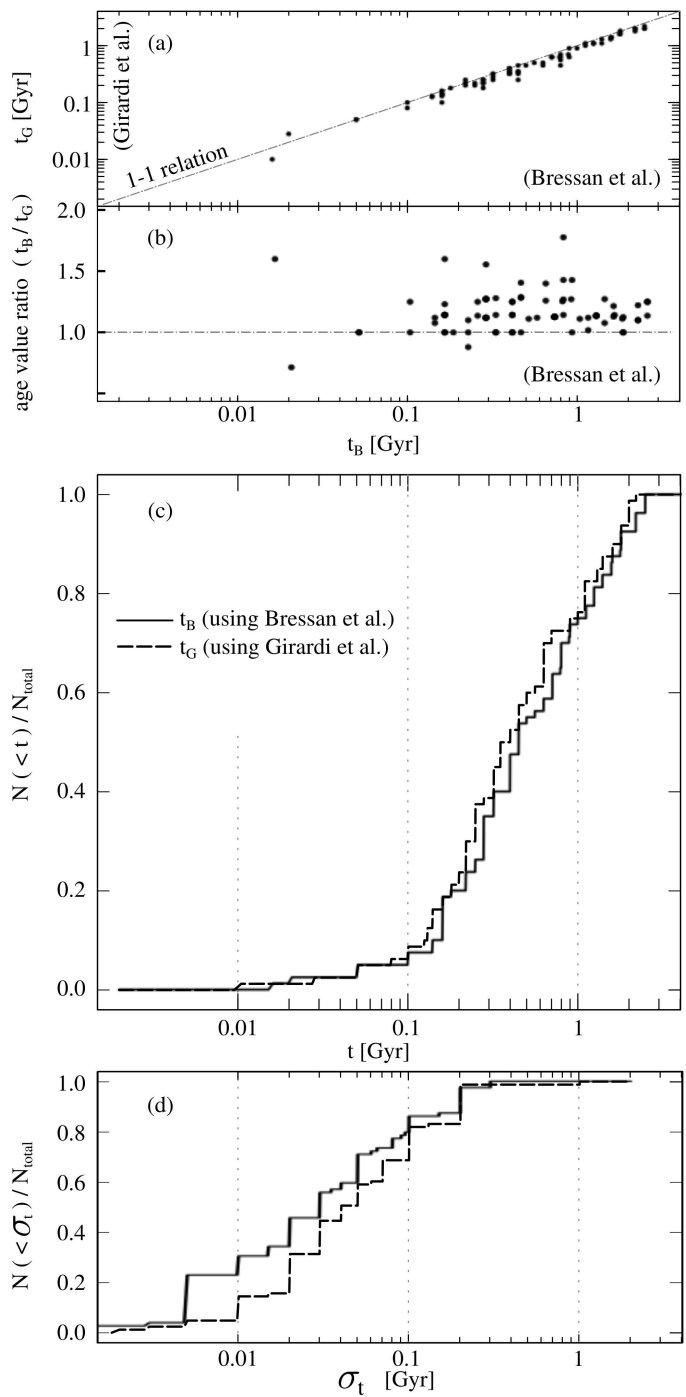}
   \caption{Age determinations using \citet{gir02} (dashed line) or \citet{bressan} (solid line) set of isochrones. In the upper panel (a), the correlation between age estimates determined using both sets of isochrones is shown.  A linear fit (not shown in the diagram) yields a slope of 0.98, which implies a slight bias towards greater age estimations when computed with Bressan et al.'s models. A 1-to-1 relation (dot dashed line) is also shown for the sake of comparison. In panel (b), the ratio between ages obtained using Bressan et al. and Girardi et al. models is presented. Values larger than 1 mean that Bressan et al. age estimates are larger. In panel (c), the age ECDFs using Bressan et al. isochrones (solid line) and Girardi et al.  isochrones (dashed line) are plotted. The uncertainty distributions are represented by the corresponding cumulative distributions in the bottom panel (d).}
   \label{f:f5} 
   \end{figure}

The ages resulting from isochrone fittings when different models of the Padova group are used will now be 
compared. We will call the ages estimated using \citet{gir02} and \citet{bressan} models $t_G$ and 
$t_B$, respectively. In order to compare the resulting ages in these two cases, we show the relation existing
between both age determinations in panels (a) and (b) of Fig. \ref{f:f5}. In panels (c) and (d) of the same figure, 
the age ECDFs and the cumulative distributions of age determination uncertainties are respectively presented. As 
shown in the upper panel (a), if a logarithmic age relation is adopted, the differences between the resulting 
ages when one or the other set of isochrones is used are not very noticeable. A linear fit in panel (a) 
yields a slope $\alpha$ = 0.98, slightly smaller than $\alpha$ = 1, which would correspond to the 
absence of a systematic differences between the two models. However, differences in some objects arise when the ratio of age determinations is considered. Panel (b) exhibits the behaviour as a function of age of the 
ratio $t_B$/$t_G$ between ages derived using these two models. Values of this ratio larger than 1 
indicate that the cluster age determinations using Bressan et al. isochrones are larger than those
using Girardi et al. isochrones. As can be seen in panel (b), $t_G$ values have been
generally underestimated compared to $t_B$ values. In fact, only in 2.5\% of the studied cluster sample can we 
find that $t_G$>$t_B$, which means that only 2 out of the 82 clusters of our sample fulfill this relation. 11 of the clusters have ages which are virtually identical whereas  in the remaining 87\% the relation 
$t_B$ > $t_G$ holds. The excess in age, measured in Gyr, is at least 12\% for 50 out of the 82 
clusters of the sample S1 and even larger than 50\% for the 5 clusters exhibiting the greatest differences.
As shown in panel (c) of Fig. \ref{f:f5}, the shapes of the two distributions in logarithmic scale are similar. The small shift between these two distributions accounts for the previously mentioned differences. 
It is seen in panel (d) that not only are $t_B$ values typically larger than those of $t_G$ but also that 
Bressan et al. age uncertainties are clearly smaller than the corresponding ones of Girardi et al. In panel (d), 
a KS-test yields p = 0.125 with $t$ in Gyr units. For example, half of the age error values 
obtained using Girardi et al. isochrones are smaller than 0.04 Gyr, while the median of the error values 
obtained from Bressan et al. isochrones is nearly 0.03 Gyr. Even if the errors involved when ages are determined 
using one or the other set of isochrones are almost the same, the ''more continuous'' distribution of the 
Bressan et al. metallicities lead to more precise fittings and hence to more accurately determined cluster 
parameters. We would like to point out that while there are global differences in age determinations, when 
individual clusters are considered, the ages inferred from the two involved models turn out to be 
practically indistinguishable, as can be observed in Table \ref{t:param}.\\

%\subsubsection{Deprojected distance distribution} 
%We finally examined the distribution of the angular deprojected distances for the LMC SCs observed in the Washington photometric system (S1 and S3 samples). According to the KS-test, it is very unlikely that the S1 and S3 distributions are random samples of the same parent distribution since the p-value is smaller than $10^{-6}$. Panel (a) of Fig. \ref{f:f6} shows that the two deprojected distance ECDFs cross each other due to their asymmetric shapes. This could be reflecting some kind of selection effect because most of the clusters we selected (S1) are projected into the LMC inner disk, while those studied in the S3 sample belong to the whole LMC disk. The deprojected distance histograms corresponding to the S3 and S1 samples are shown in panel (b) of Fig. \ref{f:f6}.

%\begin{figure}
%   \centering
 %  \includegraphics[width=\columnwidth]{f6.eps}
 %  \caption{Panel (a): cumulative deprojected distance distribution of our cluster sample S1 (continuous line histogram) 
 %  compared to that of the complementary S3 sample (dashed line histogram). Panel (b): the 
  % corresponding deprojected distance histograms are shown in the inset.}
  % \label{f:f6}
%\end{figure}

\begin{acknowledgements}
We gratefully acknowledge financial support from the Argentinian institutions CONICET, FONCYT and SECYT (Universidad Nacional de C\'ordoba). Support for T.P. is provided by the Ministry of Economy, Development, and Tourism's Millennium Science Initiative through grant IC120009, awarded to The Millennium Institute of Astrophysics, MAS. D.G. gratefully acknowledges support from the Chilean BASAL Centro de Excelencia en Astrof\'isica y Tecnolog\'ias Afines (CATA) grant PFB-06/2007. This work is based on observations made at Cerro Tololo Inter-American Observatory, which is operated by AURA, Inc., under cooperative agreement with the National Science Foundation. Part of this work was supported by the German \emph{Deut\-sche For\-schungs\-ge\-mein\-schaft, DFG\/} project number Ts~17/2--1. We especially thank the referee for his valuable comments and suggestions about the manuscript. Plots were generated using R software and post-processed with Inkscape. This research has made use of NASA's 
Astrophysics Data System.
\end{acknowledgements}

\bibliographystyle{aa}
\bibliography{paper}

\onecolumn
%\begin{longtab}
\setcounter{table}{1}
\begin{longtable}{lccccc}
%\addtocounter{table}{-2}
\caption{\label{t:clusters} LMC star clusters observed in the Washington photometric system}\\
\hline\hline
Star Cluster\,$^a$ & $\alpha_{2000}$  & $\delta_{2000}$ & l   & b & r\,$^b$ \\ 
	     &	 (hms) & (dms)   &(deg) &(deg) & (arcmin)\\
\hline
\endfirsthead
\caption{continued.}\\
\hline\hline
Star Cluster\,$^a$ & $\alpha_{2000}$ & $\delta_{2000}$ & l  & b & r\,$^b$ \\ 
	     &	 (hms) & (dms)   &(deg) &(deg) & (arcmin)\\
\hline
\endhead
\hline
HS\,8,KMHK\,5 & 04 30 40 & -66 57 25 & 278.862 & -38.343 & 0.40 \\
SL\,5,LW\,8,KMHK\,14 & 04 35 38 & -73 43 54 & 286.402 & -35.237 & 0.60 \\
NGC\,1644,SL\,9,LW\,11,ESO\,84-30,KMHK\,18 & 04 37 39 & -66 11 58 & 277.649 & -37.963 & 1.65 \\
SL\,8,LW\,13,KMHK\,21 & 04 37 52 & -69 01 42 & 280.975 & -36.949 & 0.75 \\
SL\,13,LW\,17,KMHK\,31 & 04 39 42 & -74 01 02 & 286.573 & -34.861 & 0.65 \\
KMHK\,58 & 04 43 14 & -73 48 43 & 286.224 & -34.721 & 0.43 \\
KMHK\,72 & 04 46 05 & -66 54 41 & 278.156 & -36.940 & 0.33 \\
SL\,33,LW\,59,KMHK\,91 & 04 46 25 & -72 34 06 &	284.717 & -34.986 & 0.55 \\
SL\,35,LW\,58,KMHK\,84 & 04 46 40 & -67 41 07 & 279.052 & -36.645 & 0.44 \\
KMHK\,95 & 04 47 26 & -67 39 36 & 278.994 & -36.584 & 0.23 \\
SL\,41,LW\,64,KMHK\,105 & 04 47 30 & -72 35 18 & 284.704 & -34.903 & 0.72 \\
NGC\,1697,SL\,44,ESO\,56-5,KMHK\,110 & 04 48 37 & -68 33 31 & 280.011 & -36.196 & 1.15 \\
KMHK\,123 & 04 49 00 & -72 38 24 & 284.713 & -34.780 & 0.30 \\
KMHK\,112 & 04 49 07 & -67 20 30 & 278.556 & -36.528 & 0.50 \\
KMHK\,128 & 04 49 14 & -72 03 24 & 285.177 & -34.613 & 0.26 \\
SL\,48,LW\,68,KMHK\,133 & 04 49 27 & -72 46 54 & 284.859 & -34.698 & 0.45 \\
LW\,69,KMHK\,137 & 04 49 41 & -72 14 50 & 284.243 & -34.873 & 0.28 \\
KMHK\,151 & 04 50 21 & -72 49 36 & 284.881 & -34.619 & 0.28 \\
BSDL\,77 & 04 50 29 & -67 19 36	& 278.489 & -36.407 & 0.30 \\
SL\,54,LW\,78,KMHK\,162 & 04 50 48 & -72 34 36 & 284.582 & -34.677 & 0.55 \\
BSDL\,87 & 04 50 58 & -67 36 36 & 278.808 & -36.279 & 0.25 \\
HS\,38,KMHK\,148 & 04 51 11 & -67 32 01 & 278.710 & -36.282 & 0.35 \\
HS\,41,KMHK\,158 & 04 51 30 & -67 27 15 & 278.605 & -36.276 & 0.29 \\
KMHK\,183 & 04 51 41 & -72 13 13 & 284.147 & -34.739 & 0.36 \\
SL\,73,LW\,86,KMHK\,214 & 04 52 45 & -72 31 05 & 284.454 & -34.561 & 0.34 \\
SL\,72,LW\,87,KMHK\,217 & 04 52 54 & -72 10 23 & 284.055 & -34.668 & 0.43 \\
KMHK\,229 & 04 53 52 & -69 34 14 & 281.016 & -35.435 & 0.40 \\
BSDL\,192 & 04 54 05 & -69 40 54 & 281.138 & -35.382 & 0.16 \\
BSDL\,194 & 04 54 05 & -69 45 30 & 281.227 & -35.359 & 0.21 \\
NGC\,1751,SL\,89,ESO\,56-23,KMHK\,239 & 04 54 12 & -69 48 25 & 281.280 & -35.334 & 0.75 \\
SL\,96,H88-25,KMHK\,256 & 04 55 01 & -67 42 51 & 278.795 & -35.880 & 0.41 \\
H88-26 & 04 55 03 & -67 57 52	& 279.089 & -35.806 & 0.45 \\
H88-32 & 04 55 39 & -67 43 34 & 278.788 & -35.819 & 0.21 \\
H88-34,KMHK\,285,MSX LMC\,1238 & 04 55 39 & -67 49 19 & 278.900 & -35.793 & 0.29 \\
KMHK\,286 & 04 55 42 & -67 46 54 & 278.851 & -35.799 & 0.38 \\
BSDL\,268 & 04 55 52 & -69 42 21 & 281.110 & -35.227 & 0.26 \\
BRHT\,60b,H88-41,KMHK\,309s &	04 56 26 & -67 56 19 & 279.013 & -35.689 & 0.30 \\
NGC\,1764,SL\,115,KMHK\,308 & 04 56 28 & -67 41 46 &	278.725 & -35.754	& 0.55 \\
H88-40,KMHK\,310 & 04 56 29 &	-67 37 22 & 278.638 & -35.772 & 0.38 \\
SL\,124w,KMHK\,324w & 04 56 29 & -69 59 00 & 281.413 & -35.094 & 0.29 \\
SL\,124e,KMHK\,324e & 04 56 32 & -69 58 54 & 281.409 & -35.090 & 0.29 \\
KMHK\,335 & 04 56 51 & -70 06 03 & 281.537 & -35.029 & 0.30 \\
BRHT\,45b & 04 56 52 & -68 00 20 & 279.079 & -35.632 & 0.25 \\
HS\,72,BRHT\,45a,KMHK\,326 & 04 56 54 & -68 00 08 & 279.073 & -35.630 & 0.31 \\
BSDL\,320 & 04 57 08 & -70 06 42 & 281.542 & -35.002 & 0.20 \\
SL\,126,ESO\,85-21,KMHK\,322 & 04 57 22 & -62 32 05 & 272.479 & -36.910 & 0.65 \\
SL\,132,KMHK\,348 & 04 57 26 & -67 41 07 & 278.681 & -35.668 & 0.50 \\
SL\,133,LW\,99,KMHK\,337 & 04 57 34 & -65 16 00	& 275.788 & -36.273 & 0.68 \\
H88-52,KMHK\,365 & 04 58 10 & -68 03 37 & 279.102 & -35.500 & 0.45 \\
H88-55,KMHK\,367 & 04 58 15 & -67 46 02 & 278.753 & -35.571 & 0.53 \\
BSDL\,341 & 04 58 15 & -68 02 57 & 279.086 & -35.495 & 0.28 \\
SL\,151,KMHK\,388 & 04 58 51 & -69 57 28 & 281.311 & -34.908 & 0.63 \\
H88-67 & 04 58 54 & -67 50 49 & 278.827 & -35.491 & 0.26 \\
SL\,154,H88-73,KMHK\,390 & 04 59 15 & -67 54 32 & 278.889 & -35.442 & 0.60 \\
NGC\,1793,SL\,163,ESO\,56-43,KMHK\,405 & 04 59 38 & -69 33 22 & 280.816 & -34.957 & 0.60 \\
NGC\,1795,SL\,165,ESO\,56-44,KMHK411 & 04 59 46 & -69 48 04 & 281.100 & -34.876 & 0.68 \\
SL\,162,H88-79,KMHK\,406 & 04 59 53 & -67 55 25 & 278.888 & -35.381 & 0.53 \\
BRHT\,62a,H88-84,KMHK\,412 & 05 00 04 & -67 48 02 & 278.737 & -35.396 & 0.45 \\
KMHK\,506 & 05 04 29 & -68 20 55 & 279.257 & -34.859 & 0.34 \\
BSDL\,527 & 05 04 34 & -68 12 30 & 279.088 & -34.886 & 0.21 \\
SL\,218,LOGLE\,80 & 05 05 25 & -68 30 04 & 279.411 & -34.738 & 0.46 \\
NGC\,1836,SL\,223,ESO\,56-31,BRHT\,4a & 05 05 36 & -68 37 46 & 279.557 & -34.690 & 0.73 \\
BRHT\,4b,LOGLE\,83 & 05 05 40 &	-68 38 12 & 279.563 & -34.682 & 0.48 \\
BSDL\,594, LOGLE\,87 & 05 05 54 & -67 02 58 & 277.678 &	-35.038 & 0.43 \\
NGC\,1839,SL\,226,ESO\,56-63,LOGLE\,93 & 05 06 03 & -68 37 37 & 279.542 & -34.650 & 0.80 \\
HS\,114, KMHK\,533 & 05 06 02 & -68 01 35 & 278.831 & -34.798 & 0.43 \\
NGC\,1838,SL\,225,ESO\,56-64,LOGLE\,97 & 05 06 09 & -68 26 45 &279.325 & -34.686 & 0.63 \\
HS\,116,KMHK\,536 & 05 06 13 & -68 03 53 & 278.873 & -34.772 & 0.26 \\
SL\,229,BRHT\,29a,LOGLE\,105 & 05 06 25 & -68 22 22 & 279.231 & -34.679 & 0.51 \\
SL\,230,BRHT\,29b,OGLE\,107 & 05 06 32 & -68 21 44 & 279.216 & -34.671 & 0.68 \\
BSDL\,631,LOGLE\,109 & 05 06 34 & -68 25 38 & 279.292 & -34.653 & 0.23 \\
H88-131,KMHK\,544 & 05 06 41 & -67 50 28 & 278.596 & -34.781 & 0.35 \\
LOGLE\,122 & 05 07 19 & -68 20 54 & 279.179 & -34.605 & 0.12 \\
BSDL\,654,LOGLE\,123 & 05 07 21	& -66 49 45 & 277.377 & -34.949 & 0.21 \\
LOGLE\,127 & 05 07 32 & -67 34 13 & 278.253 & -34.766 & 0.28 \\
NGC\,1846,SL\,243,ESO\,56-67,KMHK\,557 & 05 07 35 & -67 27 39 & 278.119 & -34.786 & 1.9 \\
SL\,244 & 05 07 37 & -68 32 31 & 279.399 & -34.532 & 0.50 \\
HS\,121,KMHK\,560 & 05 07 46 & -67 51 41 & 278.590 & -34.678 & 0.35 \\
BSDL\,665,LOGLE\,130 & 05 07 47	& -66 47 53 & 277.329 &	-34.914	& 0.21 \\
BSDL\,675,LOGLE\,134 & 05 07 56 & -67 21 28 & 277.990 &	-34.776	& 0.29 \\
KMHK\,575,LOGLE\,139 & 05 08 28 & -66 46 14 & 277.278 &	-34.854	& 0.47 \\
KMHK\,586 & 05 08 51 & -67 58 49 & 278.704 & -34.552 & 0.28 \\
BSDL\,716,GKK-O217 & 05 08 53 &	-68 05 01 & 278.825 & -34.525 &	0.68 \\
SL\,263,LOGLE\,144 & 05 08 54 & -66 47 08 & 277.285 & -34.809 &	0.24 \\
GKK-O222 & 05 09 00 & -67 59 00 & 278.704 & -34.537 & 0.73 \\
HS\,131 & 05 09 12 & -68 26 39 & 279.244 & -34.414 & 0.20 \\
HS\,130,KMHK\,588 & 05 09 15 & -67 42 00 & 278.362 & -34.577 & 0.28 \\
SL\,262,LW\,146,ESO\,119-40,KMHK\,582 & 05 09 21 & -62 22 46 & 271.976 & -35.577 & 0.80 \\
NGC\,1852,SL\,264,ESO\,56-71,KMHK\,594 & 05 09 23 & -67 46 42 & 278.450 & -34.547 & 0.95 \\
BSDL\,761 & 05 10 02 & -66 42 00 &	 277.155 & -34.717 & 0.32 \\
GKK-O220 & 05 10 18 & -67 51 00 & 278.512 & -34.447 & 0.78 \\
HS\,151 & 05 10 30 & -68 24 02 & 279.161 & -34.308 & 0.29 \\
BSDL\,779,LOGLE\,182	& 05 10 32 & -66 56 24 & 277.428 & -34.619 & 0.22	\\
SL\,281,KMHK\,616,LOGLE183 & 05 10 33 & -67 07 39 & 277.650  & -34.579 & 0.62 \\
SL\,290,KMHK\,628 & 05 10 36 & -70 29 15 & 281.605 & -33.806 & 0.53 \\ 
BSDL\,783,LOGLE\,186 & 05 10 39 & -66 43 45 & 277.174 & -34.651 & 0.26 \\
NGC\,1860,SL\,284,ESO\,56-75,LOGLE\,187 & 05 10 40 & -68 45 13 & 279.570 & -34.212 & 0.55 \\
BSDL794	& 05 10 46 & -67 29 06 & 278.069 &	-34.483 & 0.19	\\
H88-188,KMHK\,622,LOGLE\,191 & 05 10 54	& -67 28 16 & 278.049 & -34.474 & 0.30 \\
HS\,154,H88-189,KMHK\,625,LOGLE\,194 & 05 10 56 & -67 37 36 & 278.233 &	-34.437 &	0.35	\\
SL\,293,KMHK\,630 & 05 11 09 & -67 40 57 & 278.295 & -34.405 &	0.46	\\
HS\,156,H88-190,KMHK\,632,LOGLE\,199 & 05 11 11 & -67 37 37 & 278.227 & -34.414 & 0.25	\\
NGC\,1863,SL299,ESO\,56-77,LOGLE\,206 &	05 11 40	& -68 43 36 & 279.514 & -34.131 & 0.65 \\
SL\,300,H88-198,KMHK\,638,LOGLE\,207 & 05 11 41 & -67 33 56	& 278.142	& -34.381 & 0.46 \\
NGC\,1865,SL\,307,ESO\,56-78,LOGLE\,221 & 05 12 25	& -68 46 19 & 279.549 & -34.055 & 0.70	\\
SL\,310,KMHK\,652,LOGLE\,224 & 05 12 30 & -67 17 28	& 277.797	 & -34.359	& 0.43 \\
NGC\,1864,SL\,309,ESO\,56-79 & 05 12 40 & -67 37 24	& 278.187 & -34.276 & 0.51	\\
BSDL\,923 & 05 13 43 & -67 24 10 & 277.901 & -34.223 & 0.23	\\
HS\,178,KMHK\,667	& 05 13 48 & -66 37 12 & 276.970 &	-34.367 &	0.34	\\
NGC\,1885,SL\,338,ESO\,56-88,LOGLE\,261 & 05 15 07 & -68 58 43 & 279.729 & -33.772 & 0.65 \\
BSDL\,1024,LOGLE\,262	& 05 15 15 &	-68 52 57	& 279.614	& -33.782	&	0.41	\\
LOGLE\,264 &	05 15 21 & -69 06 27	& 279.875	 & -33.725	& 0.18 \\
H88-232 & 05 15 22	& -69 02 32 & 279.798 & -33.737 & 0.23 \\
BSDL\,1035 & 05 15 25 & -68 40 52 & 279.373 & -33.808 & 0.23 \\
HS\,198 & 05 15 26	& -69 03 02 & 279.807 &	-33.730 &	0.26	\\
HS200,LOGLE\,269	& 05 15 36 & -69 08 21 & 279.907 &	-33.697 &	0.43	\\
LOGLE\,271 &	05 15 39 & -68 54 31	& 279.635	 & -33.741	& 0.45	\\
H88-238,LOGLE\,276 & 05 15 47 & -69 14 39	 & 280.027 & -33.659 & 0.38 \\
H88-240,LOGLE\,282 & 05 16 04 &	-69 06 09	& 279.853	 & -33.664	& 0.33 \\
H88-245,LOGLE\,288 & 05 16 27 &	-69 04 49	& 279.819	& -33.635	& 0.26 \\
H88-249 & 05 16 31 & -69 10 58 & 279.939 & -33.608 & 0.23 \\
HS\,205,LOGLE\,290 & 05 16 32 & -68 55 07 &	279.627 &	-33.660 &	0.55	\\
H88-252 & 05 16 43 & -69 12 13 & 279.958  & -33.586 & 0.39 \\
H88-253,LOGLE\,296 & 05 16 50	&	-69 03 35	& 279.786 	&	-33.605	&	0.38	\\
LOGLE\,297	&	05 16 52	&	-69 04 13	&	279.798	&	-33.600	&	0.12	\\
SL\,351	&	05 16 56	&	-68 40 58	&	279.341	&	-33.673	&	0.38	\\
H88-259,LOGLE\,306	&	05 17 20	&	-69 09 25	&	279.890	&	-33.542	&	0.51	\\
H88-260,LOGLE\,307	&	05 17 20	&	-69 12 49	&	279.956	&	-33.530	&	0.36	\\
H88-261,LOGLE\,310	&	05 17 26	&	-69 06 55	&	279.838	&	-33.542	&	0.51	\\
SL\,359,KMHK\,727		&	05 17 49	&	-68 28 22	&	279.075	&	-33.635	&	0.60	\\
H88-265,LOGLE\,323	&	05 18 05	&	-69 10 18	&	279.891	&	-33.474	& 0.29	\\
H88-269,LOGLE\,337	&	05 18 41	&	-69 04 46	&	279.770	&	-33.439	&	0.33	\\
LOGLE\,340	&	05 18 47	&	-69 13 32	&	279.939	&	-33.402	&	0.32	\\
H88-270	&	05 18 47	&	-69 16 37	&	279.999	&	-33.392	&	0.23	\\
NGC\,1917,SL\,379,ESO\,56-100,LOGLE\,343 & 05 19 02 & -69 00 04 & 279.671 & -33.422 & 0.85 \\
H88-272	&	05 19 05	&	-68 52 14	& 279.515	&	-33.445	&	0.26	\\
OGLE\,346 & 05 19 09 & -69 15 36 & 279.972 & -33.363 & 0.22 \\
HS\,228 & 05 19 24 & -68 52 52 & 279.522 & -33.415 & 0.30 \\
SL\,390,LOGLE\,356	& 05 19 54 & -68 57 53 & 279.609 &	-33.354 &	0.55	\\
H88-276 & 05 19 55	& -68 48 07 & 279.418 & -33.384 & 0.26 \\
LOGLE\,363 &	05 20 04 & -69 15 55 & 279.959 & -33.282 & 0.15 \\
SL\,388,LW\,186,ESO\,85-72,KMHK\,773	& 05 20 05 & -63 28 49 & 273.090 &	-34.211 &	0.85	\\
SL\,397	&	05 20 12	&	-68 54 15	&	279.534	&	-33.341	&	0.50	\\
H88-281	&	05 20 21	&	-69 14 48	&	279.932	&	-33.262	&	0.29	\\
H88-285	&	05 21 03	&	-69 05 51	&	279.741	&	-33.229	&	0.38	\\
SL\,408A	&	05 21 05	&	-69 04 16	&	278.896	&	-34.889	&	0.41	\\
H88-286	&	05 21 07	&	-69 08 09	&	279.787	&	-33.216	&	0.20	\\
H88-287 & 05 21 09 & -69 07 02 & 279.763 & -33.216 & 0.24 \\
BSDL\,1334, \,88-259 ,LOGLE\,306 & 05 21 14 & -68 47 00 & 279.369 & -33.271 & 0.23 \\
HS\,247	&	05 21 45	&	-68 55 02	&	279.516	&	-33.201	&	0.38	\\
BSDL\,1364	&	05 21 46	&	-68 43 53		&	279.298	&	-33.233	&	0.21	\\
HS\,253,LOGLE\,403 &	05 22 03	&	-70 02 44	&	280.833	&	-32.962	&	0.29	\\
KMK88-52	&	05 22 17	&	-70 02 00	&	280.814	&	-32.945	&	0.25	\\
IC\,2134,SL\,437,LW\,198,ESO\,33-19,KMHK\,864 & 05 23 06	& -75 26 48 & 287.049 & -31.698 & 0.70	\\
HS\,264,KMHK\,845 & 05 23 12 & -70 46 40 & 281.666 & -32.725 & 0.40 \\
SL\,451,LW\,206,KMHK\,883	&	05 24 13	&	-75 34 00	&	287.166	&	-31.600	&	0.50	\\
SL\,446A,KMHK\,858	&	05 24 28	&	-67 43 43	&	278.067	&	-33.156	&	0.46	\\
SL\,444,KMHK\,861	&	05 24 30	&	-67 40 41	&	278.006	&	-33.161	&	0.54	\\
LW\,211,KMHK\,901	&	05 25 27	&	-73 34 13	&	284.858	&	-31.979	&	0.33	\\
SL\,460,LOGLE\,456	  &	05 25 28	&	-69 46 32	&	280.453	&	-32.724	&	0.38	\\
SL\,469,LOGLE\,467	 &	05 25 57	&	-69 45 04	&	280.415	&	-32.687	&	0.43	\\
KMHK\,907	 &	05 26 12	&	-70 58 53	&	281.847	&	-32.445	&	0.26	\\
BSDL\,1723,LOGLE\,473	&	05 26 24	&	-69 43 51	&	280.384	&	-32.652	&	0.33	\\
NGC\,1969,SL\,479,ESO\,56-124	&	05 26 34	&	-69 50 27	&	280.509	&	-32.619	&	0.60	\\
NGC\,1971,SL\,481,ESO\,56-128	&	05 26 46	&	-69 51 03	&	280.518	&	-32.601	&	0.51	\\
NGC\,1972,SL\,480,ESO\,56-129	&	05 26 49	&	-69 50 17	&	280.502	&	-32.599	&	0.42	\\
KMK88-57,LOGLE\,483	&	05 26 53	&	-69 48 54	&	280.473	&	-32.596	&	0.29  \\
SL\,490,LW\,217,KMHK\,939	&	05 27 18	&	-73 40 48	&	284.951	&	-31.828	&	0.58	\\
H\,14,SL\,506,LW\,220,KMHK\,967	& 05 28 39	&	-73 37 49	&	284.871	&	-31.745	&	0.70	\\
SL\,505,KMHK\,960	&	05 28 50	&	-71 37 58	&	282.560	&	-32.120 &	 0.58	\\
SL\,510,KMHK\,968	&	05 29 20	&	-70 34 46	&	281.325	&	-32.262	&	0.36	\\
KMHK\,979,GKK-O101	&	05 29 39	&	-70 59 02	&	281.790	&	-32.168	&	0.30	\\
HS\,329,KMHK\,984	&	05 29 46	&	-71 00 02	&	281.807	&	-32.156	&	0.35	\\
SL\,509,LW\,221,ESO\,85-91,KMHK\,957	& 05 29 48 & -63 38 58 & 273.152 & -33.118 & 0.93	\\
LW\,224 & 05 29 56	& -72 03 17 & 283.030 & -31.959 & 0.21 \\
KMHK\,975  &	05 29 59	& -67 52 44 & 278.151 &	-32.618 &	0.19	\\
LW\,231,KMHK\,1031 & 05 30 26 & -75 20 57 & 286.813 & -31.270 & 0.45 \\
LW\,231,KMHK\,1031 & 05 30 26 & -75 20 54 & 286.813 & -31.270 & 0.45 \\
NGC\,1997,SL\,520,LW\,226,ESO\,86-1,KMHK\,978 & 05 30 34 & -63 12 12 &	272.612 & -33.071 &	0.90	\\
KMHK\,993 & 05 30 34 & -68 09 27 & 278.469 & -32.526 & 0.28 \\
SL\,548,LW\,235,KMHK\,1035	 &	05 31 24	&	-72 02 33	&	282.990	&	-31.848	&	0.33	\\
SL\,555,LW\,236,KMHK\,1046	 &	05 31 42	&	-72 08 46	&	283.108	&	-31.805	&	0.70	\\
KMHK\,1023 & 05 31 46 & -68 14 08 & 278.544 & -32.405 & 0.31 \\
SL\,551,RHT\,38a,KMHK1027,GKK-O202	& 05 31 51 & -67 59 28 & 278.255 & -32.429 & 0.49	\\
KMHK\,1029	&	05 31 57	&	-67 52 43	&	278.122	&	-32.435 & 0.31	\\
BRHT\,38b,KMHK\,1032	&	05 31 58	&	-67 58 18	&	278.228	&	-32.420	&	0.35	\\
SL549,KMHK1013	&	05 32 03	&	-64 14 32	&	273.828	& -32.819	&	0.70	\\
KMHK\,1045 & 05 32 23 & -67 59 49 & 278.255 & -32.379 & 0.30 \\
KMHK\,1055 & 05 33 02 & -67 50 56 & 278.072 & -32.337 & 0.40 \\
H\,3,SL\,569,KMHK\,1065 & 05 33 20 & -68 09 08 & 278.424 & -32.272 & 0.90 \\
IC\,2140,SL\,581,LW\,241,ESO\,33-24,KMHK\,1106 & 05 33 21 & -75 22 35 & 286.800 & -31.084 & 1.15 \\
SL\,579,KMHK\,1085 &	05 34 13	&	-67 51 23	&	278.066	&	-32.224	&	0.48	\\
SL\,588,KMHK\,1101 &	05 34 39	&	-68 18 20	&	278.587	&	-32.129	&	0.68	\\
IC\,2146,SL\,632,LW\,258,ESO\,33-26,KMHK\,1178 & 05 37 46 & -74 46 58 & 286.058 & -30.917 & 1.65 \\
LW\,263,KMHK\,1208 & 05 39 08 & -74 51 12 & 286.120 & -30.817 & 0.48 \\
H88-306 & 05 40 24	& -69 15 10 & 279.623 & -31.505 & 0.28 \\
H88-313 & 05 41 21	& -69 03 46 & 279.391 & -31.441 & 0.20 \\
HS\,390,MHK1239	&	05 41 30	&	-69 11 06	&	279.532	&	-31.416	&	0.33	\\
H88-315,GKK-O164	&	05 41 38	&	-69 18 48	&	279.680	&	-31.391	&	0.25	\\
NGC\,2093,SL\,657,ESO\,56-23 &	05 41 49	&	-68 55 15	&	279.221	&	-31.414	& 0.70 \\
H88-320,MHK\,1248	&	05 41 58	&	-69 02 51	&	279.365	&	-31.388	&	0.30	\\
H88-321	&	05 42 08	&	-69 22 00	&	279.737	&	-31.341	&	0.19	\\
SL\,663,LW\,273,ESO\,86-22,KMHK\,1250 & 05 42 29 & -65 21 46 & 275.053 & -31.629 & 0.95 \\
H88-325 & 05 43 15 & -69 02 03 & 279.337 & -31.275 & 0.23 \\
SL\,674,ESO\,86-26,KMHK\,1281	&	05 43 20	&	-66 15 44	&	276.100	&	-31.487	&	0.80	\\
H88-326	&	05 43 29	&	-69 09 44	&	279.484	&	-31.242	&	0.24	\\
SL\,678,KMHK1283	&	05 43 35	&	-66 12 31	&	276.034	&	-31.464	&	0.58	\\
H88-327,KMHK1295	 &	05 43 38	&	-69 15 51	&	279.603	&	-31.219	&	0.35	 \\
H88-329,KMHK1297	 &	05 43 43	&	-69 13 23	&	279.554	&	-31.216	&	0.29	\\
NGC\,2108,SL\,686,ESO\,57-33,KMHK\,1304 & 05 43 56 & -69 10 50 & 279.503 & -31.200 & 0.90 \\
H88-331,MHK1313	&	05 44 11	&	-69 20 00	&	279.677	&	-31.165	&	0.31	\\
SL\,691,BRHT\,40a,KMHK\,1319 & 05 44 14 & -70 39 20	& 281.213 & -31.027 & 0.38 \\
SL\,692,BRHT\,40b,KMHK\,1320 & 05 44 15 &	-70 40 10 & 281.229 & -31.025	& 0.48 \\
BSDL\,2938,LOGLE\,717	&	05 44 42	&	-70 25 31	&	280.941	&	-31.013	&	0.23	\\
HS\,406,KMHK\,1332,LOGLE\,720	&	05 44 47	&	-70 24 22	&	280.917	&	-31.008	&	0.33	\\
HS\,409,KMHK\,1336,LOGLE\,721	&	05 44 57	&	-70 19 59	&	280.831	&	-31.001	&	0.28	\\
BSDL\,2950,LOGLE\,723	&	05 45 01	&	-70 32 34	&	281.074	&	-30.974	&	0.23	\\
BSDL\,2963,LOGLE\,727	&	05 45 20	&	-70 36 06	&	281.139	&	-30.941	&	0.23	\\
SL\,704,KMHK\,1343,LOGLE\,728	&	05 45 25	&	-70 24 05	&	280.905	&	-30.955	&	0.39	\\
H88-333	&	05 45 27	&	-69 20 43	&	279.679	&	-31.052	&	0.26	\\
HS\,410,KMHK\,1344,LOGLE\,729	&	05 45 32	&	-70 45 34	&	281.320	&	-30.910	&	0.36	\\
BSDL\,2972,LOGLE\,731	&	05 45 46	&	-70 43 09	&	281.271	&	-30.894	&	0.24	\\
HS411,MHK1345	&	05 45 50	&	-69 22 49	&	279.716	&	-31.016	&	0.23	\\
HS412,MHK1347	&	05 45 56	&	-69 16 19	&	279.589	&	-31.016	&	0.35	\\
BSDL\,2978,LOGLE\,732	&	05 45 59	&	-70 43 46	&	281.281	&	-30.876 &	0.20	\\
LOGLE\,733	&	05 46 11	&	-70 43 12	&	281.268	&	-30.860	&	0.17	\\
SL\,707,KMHK\,1353 & 05 46 12 & -69 04 57 & 279.366 & -31.008 & 0.65 \\
BSDL\,2993,LOGLE\,735	&	05 46 37	&	-70 46 33	&	281.329	&	-30.820	&	0.24	\\
HS\,414,BRHT\,42b,KMHK\,1365	&	05 46 41	& -70 50 52	&	281.411	&	-30.807	&	0.44	\\
SL\,716,BRHT\,42a,KMHK\,1367	&	05 46 47	&	-70 49 58	&	281.393	&	-30.800	&	0.53	\\
BSDL\,3001,LOGLE\,738	&	05 46 48	&	-70 35 21	&	281.111	&	-30.823	&	0.40	\\
BSDL\,2995 & 05 46 51	&	-69 25 11	&	279.754	&	-30.923	&	0.23	\\
BSDL\,3000,LOGLE\,739	& 05 46 51 & -70 30 40 &	281.019 &	-30.825 &	0.24	\\
BSDL\,3003,LOGLE\,740	& 05 46 52 & -70 48 21 &	281.361 &	-30.796 &	0.23	\\
H88-334,KMHK\,1363 & 05 46 52 & -69 11 23 & 279.486 & -30.940 & 0.36 \\
BSDL\,3050 &	05 48 00 & -70 28 30	 & 280.968 & -30.734 & 0.34 \\
KMHK\,1389 &	05 48 12 & -70 28 00 & 280.956 & -30.718 & 0.38 \\
BSDL\,3060 &	05 48 12 & -70 33 24	 & 281.061 & -30.710 & 0.37 \\
HS\,420,KMHK\,1403 &	05 48 28	&	-70 32 52	&	281.049	&	-30.688	&	0.34	\\
BSDL\,3072 &	05 48 33	&	-70 29 00	&	280.973	&	-30.687	&	0.40	\\
BSDL\,3071 & 05 48 35 & -70 18 39 & 280.773 & -30.699 & 0.20 \\
KMHK\,1408 & 05 48 46 & -70 28 23 & 280.959 & -30.670 & 0.55 \\
SL\,736 &	05 49 17 & -70 47 54	 & 281.331 & -30.599 & 0.36 \\
HS\,424,KMHK\,1425 & 05 49 36 & -70 41 35 & 281.207	& -30.582	&	0.39	\\
H\,7,SL\,735,ESO\,57-43,BM\,109 & 05 50 03 & -67 43 05	& 277.753	& -30.746 & 1.15 \\
SL\,748,KMHK\,1437 & 05 50 15 & -70 25 40 &	280.895 &	-30.550 &	0.60	\\
HS\,427,KMHK\,1443 & 05 50 17 & -70 36 56 & 281.111 & -30.532 & 0.43	\\
KMHK\,1448 & 05 50 28 & -70 32 33	& 281.027	& -30.523	& 0.34 \\
BSDL\,3123 & 05 50 45 & -70 34 34	& 281.063 & -30.497 & 0.23 \\
C\,11	& 05 50 48 & -71 42 28 & 282.371 & -30.397 & 0.20 \\
BSDL\,3158 & 05 52 11 &	-71 51 30	& 282.533	& -30.276	& 0.46 \\
KMHK\,1504 & 05 53 15 & -71 53 32	& 282.563	& -30.191	& 0.32 \\
SL\,769,KMHK1499 & 05 53 23	& -70 04 16 & 280.459 &	-30.310 &	0.90	\\
H88-365,KMHK\,1507 & 05 53 27 & -71 41 10 & 282.325 & -30.192 & 0.34 \\
SL\,775,LW\,327,KMHK\,1506 & 05 53 27	& -71 42 57 & 282.359 &	-30.189 &	0.60	\\
OHSC\,28	& 05 55 35 & -62 20 43 &	271.508 &	-30.237 &	0.35	\\
NGC\,2161,SL\,789,LW\,337,ESO\,33-31,KMHK\,1544 &	05 55 42	& -74 21 14 & 285.376 & -29.793 & 1.15 \\
NGC\,2153,SL\,792,LW\,341,ESO\,86-43,KMHK\,1555 & 05 57 51 &	-66 24 02	& 276.200	& -30.025	& 0.75 \\
NGC\,2155,SL\,803,LW\,347,ESO\,86-45,KMHK\,1563 & 05 58 33 &	-65 28 37	& 275.134	& -29.958	& 1.20 \\
SL\,817,KMHK\,1588 & 06 00 38 & -70 04 10 & 280.425 & -29.695	& 0.70 \\
SL\,826,LW\,363,KMHK,1606 & 06 01 52 & -72 21 19 & 283.046 & -29.500 & 0.75	\\
ESO\,121-03,KMHK\,1591  & 06 02 02 & -60 31 24 & 269.451 & -29.382 & 1.05 \\
LW\,393,KMHK\,1648 & 06 06 31 & -72 13 35 & 282.882 & -29.151 & 0.26 \\
LW\,397,KMHK\,1657 & 06 07 29 & -72 29 39 & 283.187 & -29.071 & 0.34 \\
SL\,842,LW\,399,ESO\,86-61,KMHK\,1652 & 06 08 15 & -62 59 15 & 272.323 & -28.814 & 0.85	\\
KMHK\,1668 & 06 08 53 & -72 23 02 & 283.056 & -28.968 & 0.29 \\
NGC\,2213,SL\,857,LW\,419,ESO\,57-70,KMHK\,1681 &	06 10 42	& 71 31 44 & 282.078 & -28.839 & 1.05	\\
SL\,862,LW\,431,ESO\,57-75,KMHK\,1692 & 06 13 27 & -70 41 45 & 281.128 & -28.613 & 0.85 \\
SL\,870,LW\,440,KMHK\,1705 & 06 14 28 & -72 36 34 & 283.310 & -28.546 & 0.58	\\
SL\,869,LW\,341,KMHK\,1704 & 06 14 41 & -69 48 07 & 280.114 & -28.490 & 0.80 \\
KMHK\,1702 & 06 14 54 & -72 30 19 & 283.190 & -28.586 & 0.31 \\
OHSC\,33,KMHK\,1714 &	06 15 17 & -73 47 07	& 284.647	& -28.482	& 0.20 \\
SL\,874,LW\,446,ESO\,57-77,KMHK\,1713 & 06 15 57 & -70 04 23 & 280.426 & -28.390 & 0.75	\\
KMHK\,1719 & 06 17 19 & -70 03 39	& 280.417	& -28.273	& 0.44 \\
LW\,469,KMHK\,1742 & 06 21 34 &	-72 47 24	& 283.522	& -28.021	& 0.53 \\
SL\,896,LW\,480,KMHK\,1758 & 06 29 58 & -69 20 01 & 279.677 & -27.131 & 0.50	\\
OHSC\,37,KMHK\,1762 & 07 07 39 & -69 59 02 & 280.983 & -24.011 & 0.29	\\
\hline
\end{longtable}
 
\tablefoot{
$(a)$ Cluster identifications from (SL): \citet{sl}; (LW): \citet{lw}; (HS): \citet{hs}; (C): \citet{h75}; H88: \citet{h88}; (OHSC): \citet{ohsc}; (KMK): \citet{kmk88}; (KMHK): \citet{kmhk}; (BRHT): \citet{brht}; (LOGLE): \citet{logle98,logle99}; (BSDL): \citet{bsdl}. $(b)$ Obtained from \citet{b08} }
%\end{longtab}
\twocolumn    

\clearpage
\newpage
\onecolumn
%\begin{longtab}
\setcounter{table}{2}
\begin{longtable}{lcccccccccc}
\caption{\label{t:param} Fundamental parameters for the star cluster sample}\\
\hline\hline
ID  & Radius & Deproj. & $E(B-V)$ & Age$_I$ &[Fe/H]$_I$ &  Age$_{II}$ &  [Fe/H]$_{II}$ & Notes & Ref	\\
      &(arcmin)&dist (deg) &	& (Gyr) & (dex) & ($\pm$ 0.30 Gyr)  &	($\pm$ 0.3 dex)  &	&	\\
\hline
\endfirsthead
\caption{continued.}\\
\hline\hline
ID  & Radius & Deproj. & $E(B-V)$ & Age$_I$ &[Fe/H]$_I$ &  Age$_{II}$ &  [Fe/H]$_{II}$ & Notes & Ref	\\
      &(arcmin)&dist (deg) &	& (Gyr) & (dex) & ($\pm$ 0.30 Gyr)  &	 ($\pm$ 0.3 dex)  &	&	\\
\hline
\endhead
\hline
HS\,8 & . . . & 6.2 & 0.026 & . . . & . . . & 1.60 & . . . & . . . &  1 \\
SL\,5 & . . . & 6.8 & 0.082 & . . . & . . . & 2.50 & . . . & . . . &  1 \\
NGC\,1644 & . . . & 6.3 & 0.018 & . . . & . . . & 2.50 & . . . & . . . &  1 \\
SL\,8 &  . . .  & 4.2 & 0.040 & . . . & . . . &  1.60  / 1.80 & -0.50 &  a  &  2,3 \\
SL\,13 & . . . & 6.8 & 0.049 & . . . & . . . & 2.50 & . . . & . . . &  1 \\
KMHK\,58 & . . . & 6.5 & 0.089 & . . . & . . . & 1.60 & . . . & . . . &  1 \\
LW\,54   & 0.30 & 5.0 & 0.000 &  0.40 $\pm$ 0.08  & -0.40 & . . . &  . . .  & a  &  4 \\
SL\,33  & 0.90 & 5.1 & 0.116 &  2.20  $_{-0.2}^{+0.3}$  & -0.40 &  . . .  & -0.60 &  m1  &  5,17 \\
SL\,35 & . . . & 4.3 & 0.051 & . . . & . . . & 1.50 & . . . & a & 1 \\
KMHK\,95   & 0.35 & 4.2 & 0.041 &  0.35 $\pm$ 0.07  & -0.40 &  . . .  &  . . .  &  . . .  &  4 \\
SL\,41  & 0.99 & 5.1 & 0.116 &  1.58 $\pm$ 0.20  & -0.57 & 1.40 &  . . .  &  . . .  &  5,17 \\
NGC\,1697  & 1.67 & 3.5 & 0.040 &  0.70 $\pm$ 0.10  & 0.00 & . . . & . . . & . . . &  6 \\
KMHK\,123  & 0.50 & 5.0 & 0.118 &  1.12 $\pm$ 0.10  & -0.54 & . . . & . . . & . . . &  5,17 \\
KMHK\,112 & . . . & 4.4 & 0.048 & . . . & . . . & 1.25 &  & a &   1 \\
KMHK\,128  & 0.50 & 5.4 & 0.111 &  1.60 $\pm$ 0.20  & -0.84 & . . . & -0.90 & . . . &  5,17 \\
SL\,48  & 0.90 & 5.1 & 0.118 &  2.50 $\pm$ 0.30  & -0.72 & 2.10 & -0.80 & . . . &  7 \\
LW\,69  & 0.54 & 4.6 & 0.122 &  1.80 $\pm$ 0.20  & -0.72 & 1.70 & . . . & . . . &  5,17 \\
KMHK\,151  & 0.77 & 5.1 & 0.118 &  1.40 $\pm$ 0.20  & -0.72 & 1.40 & -0.80 & . . . &  5,17 \\
BSDL\,77  & 0.40 & 4.3 & 0.000 &  0.79 $\pm$ 0.16  & -0.40 & . . . & . . . & . . . & 4 \\
SL\,54  & 0.90 & 4.9 & 0.120 &  1.00 $\pm$ 0.10  & -0.47 & . . . & . . . & . . . & 5,17 \\
BSDL\,87 & 0.18 & 4.0 & 0.050 &  0.08 $\pm$ 0.01  & -0.40 & . . . & . . . &  b  &  8 \\
HS\,38  & 0.25 & 4.0 & 0.050 &  0.40 $\pm$ 0.10  & -0.40 & . . . & . . . & a,b & 9 \\
HS\,41   & 0.18 & 4.1 & 0.048 &  0.06 $\pm$ 0.01  & -0.40 & . . . & . . . &  a  &  8 \\
KMHK\,183  & 0.72 & 4.5 & 0.122 &  0.79 $\pm$ 0.09  & -0.40 & . . . & . . . & a &  9 \\
SL\,73  & 0.86 & 4.7 & 0.120 &  1.78 $\pm$ 0.20  & -0.70 & 1.60 & -0.80 & . . . &  5,17 \\
SL\,72  & 0.72 & 4.4 & 0.133 &  0.28 $\pm$ 0.03  & -0.40 & . . . & . . . & . . . &  5,17 \\
KMHK\,229  & 0.25 & 2.6 & 0.100 &  1.00 $\pm$ 0.20  & -0.40 & . . . & . . . & a,b &  9 \\
BSDL\,192   & 0.18 & 2.6 & 0.102 &  0.10 $\pm$ 0.01  & -0.40 & . . . & . . . &  m2 / a,b  & 8 \\
BSDL\,194   & 0.14 & 2.6 & 0.102 &  0.25 $\pm$ 0.03  & -0.40 & . . . & . . . &  b &  8 \\
NGC\,1751 & . . . & 2.6 & 0.102 & . . . & . . . & 1.30 & . . . & . . . & 1 \\
SL\,96 & . . . & 3.6 & 0.058 & . . . & . . . & 1.60 & . . . & . . . & 1 \\
H88-26  & 0.33 & 3.3 & 0.060 &  0.80 $\pm$ 0.20  & -0.40 & . . . & . . . & b &  9 \\
H88-32 & 0.14 & 3.5 & 0.058 &  0.25 $\pm$ 0.03  & -0.40 & . . . & . . . &  b  & 8  \\
H88-34 & 0.14 & 3.1 & 0.058 &  0.25 $\pm$ 0.03  & -0.40 & . . . & . . . &  b &  8 \\
H88-33  & 0.30 & 3.4 & 0.066 &  0.16-0.32 $\pm$ 0.20  & -0.40 & . . . & . . . & b &  4 \\
BSDL\,268 & 0.99 & 2.5 & 0.102 &  0.09 $\pm$ 0.02  & -0.40 & . . . & . . . & a,b &  4 \\
BRHT\,60b   & 0.14 & 3.2 & 0.060 &  0.08 $\pm$ 0.01  & -0.40 & . . . & . . . &  m3 / a,b  & 8 \\
NGC\,1764   & 0.23 & 3.5 & 0.058 &  0.08 $\pm$ 0.01  & -0.40 & . . . & . . . &  a,b  &  8 \\
H88-40  & 0.33 & 3.5 & 0.060 &  0.70 $\pm$ 0.20  & -0.40 & . . . & . . . & b &  9 \\
SL\,124w & 0.14 & 2.5 & 0.115 &  0.50 $\pm$ 0.05  & -0.40 & . . . & . . . &  m4  & 8 \\
SL\,124e & 0.14 & 2.5 & 0.115 &  0.50 $\pm$ 0.05  & -0.40 & . . . & . . . &   m4 / b  &  8 \\
KMHK\,335   & 0.09 & 2.5 & 0.115 &  0.10 $\pm$ 0.01  & -0.40 & . . . & . . . &  b  &  8 \\
BRHT\,45b   & 0.14 & 3.2 & 0.060 &  0.08 $\pm$ 0.01  & -0.40 & . . . & . . . &   m5 / a,b  & 8 \\
BRHT\,45a & 0.45 & 3.2 & 0.076 &  0.13 $\pm$ 0.03  & -0.40 & . . . & . . . &  m5 / a,b & 4 \\
BSDL\,320   & 0.14 & 2.5 & 0.115 &  0.10 $\pm$ 0.01  & -0.40 & . . . & . . . &  a,b  &  8 \\
SL\,126 & . . . & 8.9 & 0.000 & . . . & . . . & 2.50 / 2.20 & -0.45 & . . . &  2,3 \\
SL\,132 & . . . & 3.4 & 0.057 & . . . & . . . & 1.60 & . . . & a & 1 \\
SL\,133 & 1.13 & 5.9 & 0.020 &  2.00 $\pm$ 0.20  & -0.70 & 2.30 & . . . & . . . &  6 \\
H88-52  & 0.45 & 3.0 & 0.041 &  1.12 $\pm$ 0.23  & -0.40 & 1.40 & . . . &  m6 / b  &  1,4 \\
H88-55  & 0.33 & 3.3 & 0.060 &  0.50 $\pm$ 0.10  & -0.40 & . . . & . . . & b &  9 \\
BSDL\,341 & 0.40 & 3.0 & 0.086 &  0.28 $\pm$ 0.06  & -0.40 & . . . & . . . &  m6 / b  &  4 \\
SL\,151 & . . . & 2.3 & 0.102 & . . . & . . . & 1.50 & . . . & . . . & 1 \\
H88-67 & . . . & 3.1 & 0.062 & . . . & . . . & 1.70 & . . . & . . . &  1 \\ 
SL\,154 & 0.33 & 3.0 & 0.060 &  0.50 $\pm$ 0.10  & -0.40 & . . . & . . . & b &  9 \\
NGC\,1793 & 0.42 & 2.1 & 0.107 &  0.11 $\pm$ 0.02  & -0.40 & . . . & . . . & a,b &  4 \\
NGC\,1795 & . . . & 2.1 & 0.096 & . . . & . . . & 1.60 & . . . & . . . & 1 \\
SL\,162 & . . . & 3.0 & 0.062 & . . . & . . . & 1.50 & . . . & a & 1 \\
BRHT\,62a   & 0.32 & 3.1 & 0.062 &  0.16 $\pm$ 0.02  & -0.40 & . . . & . . . &  a,b  &  8 \\
KMHK\,506  & 0.17 & 2.2 & 0.060 &  0.56 $\pm$ 0.07  & -0.40 & . . . & . . . & . . . &  9 \\
BSDL\,527 & . . . & 2.4 & 0.059 & . . . & . . . & 1.40 & . . . & . . . &  1 \\
SL\,218 & . . . & 2.0 & 0.060 &  0.05 $\pm$ 0.01  & -0.40 & . . . & . . . & a,b,c &  10 \\
NGC\,1836  & . . . & 1.9 & 0.060 &  0.40 $\pm$ 0.10  & 0.00 & . . . & . . . &  m7 / b  &  11 \\
BRHT\,4b   & . . . & 1.9 & 0.060 &  0.10 $\pm$ 0.02  & -0.40 & . . . & . . . &  m7 / a,c  &  10 \\
BSDL\,594  & 0.63 & 3.4 & 0.046 &  1.58 $_{-0.17}^{+0.20}$   & -0.47 & 1.30 & . . . & c  &  5,17 \\
HS\,113  & 0.23 & 2.2 & 0.059 &  0.20 $\pm$ 0.02  & -0.40 & . . . & . . . &  m8  &  8 \\
NGC\,1839  & . . . & 1.9 & 0.060 &  0.13 $\pm$ 0.02  & -0.40 & . . . & . . . & a,b &  10 \\
HS\,114  & . . . & 2.4 & 0.055 & . . . & . . . & 1.30 & . . . & . . . & 1 \\
NGC\,1838  & . . . & 2.0 & 0.060 &  0.10 $\pm$ 0.02  & -0.40 & . . . & . . . & a &  10 \\
HS\,116   & 0.30 & 2.4 & 0.041 &  0.35 $\pm$ 0.07  & -0.40 & . . . & . . . & . . . &  4 \\
SL\,229 & 0.33 & 2.1 & 0.060 &  0.32 $\pm$ 0.10  & -0.40 & . . . & -0.40 &  m9 / c &  4,9 \\
SL\,230   & 0.42 & 2.1 & 0.081 &  0.08 $\pm$ 0.02  & -0.40 & . . . & . . . &  m9 / a  & 4 \\
BSDL\,631 & 0.25 & 2.0 & 0.000 &  0.22 $\pm$ 0.05  & -0.40 & . . . & . . . & a,c &  4 \\
H88-131 & 0.40 & 2.6 & 0.030 &  1.00 $\pm$ 0.21  & -0.40 & . . . & . . . & . . . &  4 \\
OGLE\,122   & 0.18 & 2.0 & 0.059 &  0.40 $\pm$ 0.04  & -0.40 & . . . & . . . & c & 8 \\
BSDL\,654  & 0.34 & 3.6 & 0.033 &  0.22 $\pm$ 0.03  & -0.02 & . . . & . . . & c & 5,17 \\
LOGLE\,127  & 0.23 & 2.8 & 0.061 &  0.10 $\pm$ 0.01  & -0.40 & . . . & . . . & c & 8 \\
NGC\,1846 & . . . & 2.9 & 0.061 & . . . & . . . & 1.40 & . . . & . . . & 1 \\
SL\,244 & . . . & 1.8 & 0.060 & . . . & . . . &  1.60 / 1.30 & -0.30 & b &  2,12 \\
HS\,121 & . . . & 2.5 & 0.060 & . . . & . . . & 1.50 & . . . & . . . &   1 \\
BDSL\,665  & 0.27 & 3.6 & 0.033 &  0.90 $\pm$ 0.10  & -0.41 & . . . & . . . & c & 5,17 \\
BSDL\,675  & 0.41 & 3.0 & 0.061 &  1.40 $\pm$ 0.20  & -0.40 & 1.40 & . . . & c & 5,17 \\
KMHK\,575  & 0.59 & 4.2 & 0.043 &  0.89  $_{-0.19}^{+0.23}$  & . . . & . . . & . . . & c & 7 \\
KMHK\,586 & . . . & 2.3 & 0.055 & . . . & . . . & 1.80 & . . . & . . . &  1 \\
BSDL\,716  & 0.50 & 2.2 & 0.060 &  0.40 $\pm$ 0.10  & -0.40 & . . . & . . . & . . . &  9 \\
SL\,263  & 0.45 & 3.6 & 0.043 &  0.02  $_{-0.005}^{+0.006}$  & -0.23 & . . . & . . . & c & 7 \\
GKK-O222  & 0.14 & 2.3 & 0.055 &  1.58 $\pm$ 0.16  & -0.40 & . . . & . . . & . . . &  8 \\
HS\,131   & 0.45 & 1.8 & 0.081 &  1.26 $\pm$ 0.26  & -0.40 & . . . & . . . & . . . &  4 \\
HS\,130  & 0.41 & 2.5 & 0.061 &  0.16 $\pm$ 0.02  & -0.41 & . . . & . . . & . . . &  5,17 \\
SL\,262 & . . . & 8.6 & 0.001 & . . . & . . . & 2.10 & -0.55 & . . . &  2,3 \\
NGC\,1852 & . . . & 2.5 & 0.060 & . . . & . . . & 1.40 & . . . & . . . & 1 \\
BSDL\,761  & 0.41 & 3.6 & 0.036 &  0.16 $\pm$ 0.02  & -0.40 & . . . & . . . & . . . &  5,17 \\
GKK-O220  & 0.23 & 2.3 & 0.060 &  0.79 $\pm$ 0.08  & -0.40 & . . . & . . . & . . . &  8 \\
HS\,151 & 0.33 & 1.8 & 0.060 &  0.79 $\pm$ 0.10  & -0.40 & . . . & . . . & . . . &  9 \\
BSDL\,779  & 0.36 & 3.3 & 0.043 &  0.10 $\pm$ 0.01  & -0.02 & . . . & . . . & c & 5,17 \\
SL\,281  & 0.95 & 3.1 & 0.052 &  0.05  $_{-0.005}^{+0.006}$  & -0.31 & . . . & . . . & a,c &  17 \\
SL\,290 & . . . & 1.7 & 0.101 & . . . & . . . & 1.20 & . . . & a & 1 \\
BSDL\,783  & 0.32 & 3.6 & 0.072 &  0.16 $\pm$ 0.01  & -0.30 & . . . & . . . & a,c &  17 \\
NGC\,1860  & . . . & 1.4 & 0.080 &  0.25 $\pm$ 0.50  & 0.00 & . . . & . . . & b,c & 11 \\
BSDL\,794  & 0.45 & 2.7 & 0.060 &  0.80  $_{-0.17}^{+0.3}$  & -0.41 & . . . & . . . &  m10  &  7 \\
H88-188  & 0.41 & 2.7 & 0.061 &  0.63  $_{-0.07}^{+0.08}$  & -0.27 & . . . & . . . &  m10 / c  & 9,17 \\
HS\,154 & 0.33 & 2.5 & 0.060 &  0.45-0.50 $\pm$ 0.10  & -0.40 & . . . & . . . &  m11 / c &  4,9 \\
SL\,293  &  0.63/0.5  & 2.5 & 0.061 &  0.40$\pm$ 0.05  &  -0.31  & . . . & . . . & . . . &  9,17 \\
HS\,156  & 0.54 & 2.5 & 0.061 &  1.25  $_{-0.1}^{+0.2}$  &  $_{-0.47}^{0.40}$  & . . . & . . . &  m11 / c  &  4,5,8,17 \\
NGC\,1863  & . . . & 1.4 & 0.060 &  0.04 $\pm$ 0.01  & -0.40 & . . . & . . . & a,b,c &  10 \\
SL\,300 & 0.67 & 2.6 & 0.060 &  0.40 $\pm$ 0.10  & -0.40 & . . . & . . . & c &  9 \\
NGC\,1865  & . . . & 1.3 & 0.060 &  0.50 $\pm$ 0.10  & -0.20 & . . . & . . . &  m12 / c  &  2,11 \\
SL\,310  & 0.68 & 2.8 & 0.061 &  0.05 $\pm$ 0.01  & -0.30 & . . . & . . . & a,c & 17 \\
NGC\,1864  & 0.99 & 2.5 & 0.061 &  0.25 $\pm$ 0.03  & -0.40 & . . . & . . . & a &  17 \\
BSDL\,923  & 0.41 & 2.7 & 0.060 &  0.10$_{-0.01}^{+0.02}$  & -0.36 & . . . & . . . & a &  17 \\
HS\,178  & 0.54 & 3.5 & 0.036 &  0.71 $\pm$ 0.08  & -0.41 & . . . & . . . & . . . & 5 \\
NGC\,1885   & 0.32 & 0.9 & 0.081 &  0.06 $\pm$ 0.01  & -0.40 & . . . & . . . &  c  & 8 \\
BSDL\,1024 & 1.00 & 1.0 & 0.080 &  0.16 $\pm$ 0.03  & -0.40 & . . . & . . . & c & 9 \\
OGLE\,264 & 0.18 & 0.7 & 0.078 &  0.63 $\pm$ 0.06  & -0.40 & . . . & . . . & c &  8 \\
H88-232   & 0.18 & 0.8 & 0.078 &  0.13 $\pm$ 0.01  & -0.40 & . . . & . . . &  m13 / c  & 8 \\
BSDL\,1035 & 0.17 & 1.2 & 0.060 &  0.50 $\pm$ 0.10  & -0.40 & . . . & . . . & . . . &  9 \\
HS\,198  & 0.14 & 0.8 & 0.078 &  0.14 $\pm$ 0.01  & -0.40 & . . . & . . . &  m13 / b,c  &  8 \\
HS\,200  & 0.18 & 0.7 & 0.086 &  0.50 $\pm$ 0.05  & -0.40 & . . . & . . . &  b,c  & 8 \\
OGLE\,271   & 0.27 & 0.9 & 0.078 &  0.40 $\pm$ 0.04  & -0.40 & . . . & . . . & c & 8 \\
H88-238   & 0.23 & 0.6 & 0.090 &  0.13 $\pm$ 0.01  & -0.40 & . . . & . . . &  b,c  & 8 \\
H88-240   & 0.27 & 0.7 & 0.081 &  0.20 $\pm$ 0.02  & -0.40 & . . . & . . . &  b,c  & 8 \\
H88-245 & 0.50 & 0.7 & 0.080 &  0.16 $\pm$ 0.04  & -0.40 & . . . & . . . & b,c &  9 \\
H88-249   & 0.14 & 0.6 & 0.090 &  0.32 $\pm$ 0.03  & -0.40 & . . . & . . . & . . . &  8 \\
HS\,205  & 0.27 & 0.8 & 0.081 &  0.10 $\pm$ 0.01  & -0.40 & . . . & . . . & c & 8 \\
H88-252   & 0.23 & 0.6 & 0.090 &  0.25 $\pm$ 0.03  & -0.40 & . . . &    &  b  &  8 \\
H88-253   & 0.18 & 0.7 & 0.081 &  0.20 $\pm$ 0.02  & -0.40 & . . . & . . . &  m14 / c & 8 \\
OGLE\,297   & 0.27 & 0.7 & 0.081 &  0.08 $\pm$ 0.01  & -0.40 & . . . & . . . &  b,c  & 8 \\
SL\,351 & 0.33 & 1.1 & 0.060 &  0.50 $\pm$ 0.10  & -0.40 & . . . & . . . & . . . &  9 \\
H88-259   & 0.18 & 0.6 & 0.081 &  0.79 $\pm$ 0.08  & -0.40 & . . . & . . . & c &  8 \\
H88-260   & 0.18 & 0.5 & 0.090 &  0.40 $\pm$ 0.04  & -0.40 & . . . & . . . & c &  8 \\
H88-261   & 0.32 & 0.6 & 0.081 &  0.50 $\pm$ 0.05  & -0.40 & . . . & . . . & c &  8 \\
SL\,359 & . . . & 1.3 & 0.060 &  . . .  & . . . & 1.60 & -0.4 & . . . &  2,12 \\
H88-265 & 0.33 & 0.5 & 0.051 &  0.20 $\pm$ 0.04  & -0.40 & . . . & . . . & b,c & 4 \\
H88-269 & 0.33 & 0.5 & 0.051 &  0.79 $\pm$ 0.16  & -0.40 & . . . & . . . & c &  4 \\
OGLE\,340   & 0.36 & 0.4 & 0.090 &  0.04 $\pm$ 0.00  & -0.40 & . . . & . . . & c &  8 \\
H88-270   & 0.18 & 0.3 & 0.090 &  1.26 $\pm$ 0.13  & -0.40 & . . . & . . . &  b  &  8 \\
NGC\,1917 & . . . & 0.6 & 0.081 & . . . & . . . & 1.30 & . . . & . . . & 1 \\
H88-272   & 0.23 & 0.8 & 0.060 &  0.32 $\pm$ 0.03  & -0.40 & . . . & . . . &  m15  & 8 \\
OGLE\,346   & 0.25 & 0.3 & 0.090 &  0.16 $\pm$ 0.02  & -0.40 & . . . & . . . & c & 8 \\
HS\,228  & 0.27 & 0.7 & 0.060 &  1.26 $\pm$ 0.13  & -0.40 & . . . & . . . &  m15 / b  & 8 \\
SL\,390  & 0.32 & 0.6 & 0.081 &  1.00 $\pm$ 0.10  & -0.40 & . . . & . . . &  b,c  &  8 \\
H88-276   & 0.23 & 0.8 & 0.060 &  0.79 $\pm$ 0.08  & -0.40 & . . . & . . . &  b  &  8 \\
OGLE\,363   & 0.18 & 0.3 & 0.090 &  0.10 $\pm$ 0.01  & -0.40 & . . . & . . . & c & 8 \\
SL\,388 & 0.50 & 7.0 & 0.030 & . . . & . . . & 2.2 / 2.6 & -0.65 & . . . & 2,3 \\
SL\,397   & 0.42 & 0.7 & 0.081 &  0.16 $\pm$ 0.03  & -0.40 & . . . & . . . & a,b &  4 \\
H88-281   & 0.23 & 0.3 & 0.090 &  0.16 $\pm$ 0.02  & -0.40 & . . . & . . . &  b  &  8 \\
H88-285   & 0.18 & 0.4 & 0.081 &  0.32 $\pm$ 0.03  & -0.40 & . . . & . . . &  m16 / b  & 8 \\
SL\,408A & 0.23 & 0.5 & 0.055 &  0.14 $\pm$ 0.01  & -0.40 & . . . & . . . &  a,b  &  8 \\
H88-286   & 0.14 & 0.4 & 0.081 &  0.32 $\pm$ 0.03  & -0.40 & . . . & . . . &  m16 / a  & 8 \\
H88-287   & 0.23 & 0.4 & 0.081 &  0.16 $\pm$ 0.02  & -0.40 & . . . & . . . &  m16 / b  &  8 \\
BSDL\,1334  & 0.18 & 0.8 & 0.060 &  1.00 $\pm$ 0.10  & -0.40 & . . . & . . . & b,c &  8 \\
HS\,247   & 0.33 & 0.6 & 0.127 &  0.35 $\pm$ 0.07  & -0.40 & . . . & . . . & b &  4 \\
BSDL\,1364  & 0.09 & 0.9 & 0.060 &  0.16 $\pm$ 0.02  & -0.40 & . . . & . . . & . . . & 8 \\
HS\,253  & 0.45 & 0.7 & 0.085 &  0.22  $_{-0.02}^{+0.03}$  & -0.36 & . . . & . . . & b,c & 17 \\
KMK88-52  & 0.45 & 0.6 & 0.085 &  0.18 $\pm$ 0.02  & +0.02 & . . . & . . . &  m17  &  17 \\
IC\,2134   & 0.50 & 6.9 & 0.107 & . . . & . . . & 1.00 & . . . & . . . &  3 \\
HS\,264 & . . . & 1.5 & 0.083 & . . . & . . . & 1.60 & . . . & a &  1 \\
SL\,451 & . . . & 7.0 & 0.106 & . . . & . . . & 2.20 & -0.70 & . . . &  2,3 \\
SL\,446A  & . . . & 2.0 & 0.060 &   2.20$\pm$ 0.50  & -0.90 &  2.30 / 2.40 & -0.75 & a &   2,12 \\
SL\,444 & . . . & 2.0 & 0.060 &  0.50 $\pm$ 0.10  & -0.40 & . . . & . . . & a &  11 \\
LW\,211  & 0.72 & 4.7 & 0.097 &  1.80  $_{-0.2}^{+0.4}$  & -0.67 & . . . & . . . & . . . & 5,17 \\
SL\,460  & 0.72 & 0.5 & 0.062 &  0.02  $_{-0.006}^{+0.012}$  & -0.47 & . . . & . . . &  m18 / c &  17 \\
SL\,469  & 0.54 & 0.5 & 0.062 &  0.16 $\pm$ 0.02  & -0.44 & . . . & . . . & c &  17 \\
KMHK\,907 & 0.35 & 1.7 & 0.091 &  0.25 $\pm$ 0.05  & -0.40 & . . . & . . . & . . . & 4 \\
BSDL\,1723  & 0.45 & 0.5 & 0.062 &  0.28 $\pm$ 0.03  & -0.40 & . . . & . . . & c &  17 \\
NGC\,1969  & 0.63 & 0.5 & 0.062 &  0.16  $_{-0.03}^{+0.09}$  & -0.54 & . . . & . . . &  m19  &  17 \\
NGC\,1971  & 0.50 & 0.5 & 0.062 &  0.14  $_{-0.3}^{+0.11}$  & -0.47 & . . . & . . . &  m19  &  17 \\
NGC\,1972  & 0.36 & 0.5 & 0.062 &  0.16  $_{-0.03}^{+0.09}$  & -0.44 & . . . & . . . &  m19  &  17 \\
KMK88-57  & 0.41 & 0.6 & 0.062 &  0.80  $_{-0.24}^{+0.2}$  & -0.54 & . . . & . . . &  m20 / c &  17 \\
SL\,490  & 0.99 & 4.8 & 0.120 &  2.20  $_{-0.40}^{+0.30}$  &  -0.66  & 1.80 & -0.80 & . . . & 7,13 \\
SL\,506  & 1.35 & 4.8 & 0.106 &  1.80 $\pm$ 0.20  & -0.66 & 1.70 & -0.80 & . . . &  17 \\
SL\,505 & . . . & 2.5 & 0.070 &   0.90$\pm$ 0.20  & -0.50 &  1.60 / 1.50 & -0.70 & a &  2,12 \\
SL\,510 & 0.17 & 1.4 & 0.080 &  0.13 $\pm$ 0.03  & -0.40 & . . . & . . . & a &  9 \\
KMHK\,979 & 0.33 & 1.8 & 0.086 &  0.08 $\pm$ 0.02  & -0.40 & . . . & . . . &  m21 \ a  &  4 \\
HS\,329   & 0.65 & 1.8 & 0.000 &  0.79-1.00 $\pm$ 0.23  & -0.40 & 1.80 & . . . &  m21  &  1,6 \\
SL\,509 & . . . & 6.6 & 0.030 & . . . & . . . &  1.40 / 1.20 & -0.85 & . . . &  2,3 \\
LW\,224 & . . . & 3.0 & 0.060 &  0.70 $\pm$ 0.10  & 0.00 & . . . & . . . &  m22  &  11 \\
KMHK\,975 & 0.35 & 1.9 & 0.051 &  0.20 $\pm$ 0.04  & -0.40 & . . . & . . . & . . . &  4 \\
LW\,231  & 0.52 & 6.7 & 0.110 &  0.80  $_{-0.17}^{+0.20}$  &  -0.50  & . . . & . . . & . . . &  7,13 \\
NGC\,1997  & 0.80 & 7.1 & 0.040 &  2.60 $\pm$ 0.50  & -0.70 & 2.70 & . . . & . . . &  6 \\
KMHK\,993   & 0.18 & 1.6 & 0.062 &  0.10 $\pm$ 0.01  & -0.40 & . . . & . . . & . . . &  8 \\
SL\,548 & . . . & 3.0 & 0.080 &  0.40 $\pm$ 0.10  & 0.00 & . . . & . . . & . . . & 11 \\
SL\,555 & . . . & 3.1 & 0.070 &  1.60 $\pm$ 0.30  & -0.70 &  1.60 / 1.80 & -0.75 & a &  2,12 \\
KMHK\,1023 & . . . & 1.6 & 0.062 & . . . & . . . & 1.70 & . . . & . . . &   1 \\
SL\,551   & 0.33 & 1.8 & 0.091 &  0.14 $\pm$ 0.03  & -0.40 & . . . & . . . &  m23 / a,b &  4 \\
KMHK\,1029  & 0.27 & 2.0 & 0.058 &  0.10 $\pm$ 0.01  & -0.40 & . . . & . . . &  b  &  8 \\
BRHT\,38b & 0.50 & 1.9 & 0.081 &  0.18 $\pm$ 0.04  & -0.40 & . . . & . . . &  m23 / b  &  4 \\
SL\,549 & . . . & 5.9 & 0.040 &  2.00 $\pm$ 0.50  & -0.90 &  1.30 / 2.00 & . . . & . . . &  2,12 \\
KMHK\,1045 & 0.17 & 1.9 & 0.060 &  0.60 $\pm$ 0.10  & -0.40 & . . . & . . . & b & 9 \\
KMHK\,1055 & 0.17 & 2.0 & 0.060 &  1.00 $\pm$ 0.20  & -0.40 &  . . .  &  . . .  &  a,b  &  9 \\
H\,3 & . . . & 1.8 & 0.062 & . . . & . . . & 2.50 & . . . & . . . & 1 \\
IC\,2140  & 1.17 & 6.8 & 0.111 &  2.50  $_{-0.5}^{+0.6}$  & -0.84 & 2.10 & -1.10 & . . . & 7 \\
SL\,579   & 0.30 & 2.1 & 0.066 &  0.14 $\pm$ 0.03  & -0.40 & . . . & . . . & a,b &  4 \\
SL\,588 & 0.67 & 1.7 & 0.060 &  0.40 $\pm$ 0.10  & -0.40 & . . . & . . . & b &  9 \\
IC\,2146 & . . . & 6.1 & 0.117 & . . . & . . . & 1.60 & . . . & . . . & 1 \\
LW\,263 & . . . & 6.2 & 0.117 & . . . & . . . & 1.80 & . . . & . . . & 1 \\
H88-306   & 0.14 & 1.8 & 0.063 &  0.13 $\pm$ 0.01  & -0.40 & . . . & . . . &  m24  & 8 \\
H88-313   & 0.18 & 1.9 & 0.064 &  0.13 $\pm$ 0.01  & -0.40 & . . . & . . . &  b  &  8 \\
HS\,390   & 0.33 & 1.9 & 0.228 &  0.18 $\pm$ 0.04  & -0.40 & . . . & . . . & b &  4 \\
H88-315 & 0.18 & 2.0 & 0.063 &  0.08 $\pm$ 0.01  & -0.40 & . . . & . . . &  a,b  &  8 \\
NGC\,2093  & 0.50 & 2.0 & 0.070 &  0.25 $\pm$ 0.05  & -0.40 & . . . & . . . &  m25 / a,b  &  9 \\
H88-320 & 0.45 & 2.0 & 0.168 &  0.16 $\pm$ 0.03  & -0.40 & . . . & . . . & a,b &  4 \\
H88-321   & 0.18 & 2.0 & 0.063 &  0.08 $\pm$ 0.01  & -0.40 & . . . & . . . &  b  &  8 \\
SL\,663 & 1.93 & 4.8 & 0.040 &  2.80 $\pm$ 0.35  & -0.70 & 3.30 & . . . & . . . &  6 \\
H88-325   & 0.18 & 2.1 & 0.072 &  0.22 $\pm$ 0.02  & -0.40 & . . . & . . . & . . . &  8 \\
SL\,674 & . . . & 3.9 & 0.050 &   2.00 $\pm$ 0.40  & -0.90 &  2.10 / 2.30 & -0.80 & a &  2,12 \\
H88-326   & 0.18 & 2.1 & 0.072 &  0.32 $\pm$ 0.03  & -0.40 & . . . & . . . &  a,b  &  8 \\
SL\,678 & . . . & 4.0 & 0.050 &  1.50 $\pm$ 0.30  & -0.80 & 2.00 & -0.80 & a &  12 \\
H88-327   & 0.27 & 2.1 & 0.074 &  0.03 $\pm$ 0.00  & -0.40 & . . . & . . . &  m26  &  8 \\
H88-329  & 0.18 & 2.2 & 0.072 &  0.06 $\pm$ 0.01  & -0.40 & . . . & . . . &  b  &  8 \\
NGC\,2108 & . . . & 2.2 & 0.072 & . . . & . . . & 1.25 & . . . & . . . & 1 \\
H88-331 & 0.40 & 2.2 & 0.117 &  0.50 $\pm$ 0.10  & -0.40 & . . . & . . . & . . . &  4 \\
SL\,691  & 0.38 & 2.3 & 0.068 &  0.28  $_{-0.08}^{+0.07}$  & -0.47 & . . . & . . . &  m27 / a,c &  17 \\
SL\,692  & 0.47 & 2.3 & 0.068 &  0.25 $\pm$ 0.03  & -0.41 & . . . & . . . &  m27 / a,c &  17 \\
BSDL\,2938  & 0.45 & 2.3 & 0.067 &  0.45  $_{-0.10}^{+0.18}$  & -0.23 & . . . & . . . & . . . &  17 \\
HS\,406  & 0.32 & 2.3 & 0.067 &  0.32  $_{-0.04}^{+0.03}$  & -0.23 & . . . & . . . & c &  17 \\
HS\,409  & 0.50 & 2.3 & 0.067 &  0.45  $_{-0.13}^{+0.11}$  & -0.27 & . . . & . . . & c &  7 \\
BSDL\,2950  & 0.27 & 2.3 & 0.067 &  0.71 $\pm$ 0.08  & -0.36 & . . . & . . . & c &  17 \\
BSDL\,2963  & 0.52 & 2.4 & 0.068 &  1.25  $_{-0.1}^{+0.2}$  & 0.00 & . . . & . . . & c &  17 \\
SL\,704  & 0.45 & 2.3 & 0.067 &  0.45  $_{-0.05}^{+0.11}$  & -0.16 & . . . & . . . & c &  17 \\
H88-333 & 0.33 & 2.3 & 0.070 &  0.40 $\pm$ 0.10  & -0.40 & . . . & . . . & . . . &  9 \\
HS\,410  & 0.63 & 2.5 & 0.068 &  0.56 $\pm$ 0.06  & -0.27 & . . . & . . . & c &  17 \\
BSDL\,2972  & 0.43 & 2.5 & 0.068 &  0.71  $_{-0.1}^{+0.09}$  & 0.00 & . . . & . . . & c &  17 \\
HS\,411   & 0.25 & 2.4 & 0.173 &  0.28 $\pm$ 0.06  & -0.40 & . . . & . . . & . . . &  4 \\
HS\,412   & 0.42 & 2.4 & 0.173 &  0.13 $\pm$ 0.03  & -0.40 & . . . & . . . & a &  4 \\
BSDL\,2978  & 0.27 & 2.5 & 0.068 &  0.90  $_{-0.27}^{+0.2}$  & -0.23 & . . . & . . . & c &  17 \\
LOGLE\,733  & 0.27 & 2.5 & 0.068 &  0.80  $_{-0.24}^{+0.2}$  & -0.10 & . . . & . . . & c & 17 \\
SL\,707 & . . . & 2.4 & 0.072 & . . . & . . . & 2.30 & . . . & . . . & 1 \\
BSDL\,2993  & 0.59 & 2.5 & 0.068 &  0.71 $\pm$ 0.08  & -0.36 & . . . & . . . & c &  17 \\
HS\,414  & 0.45 & 2.6 & 0.086 &  0.32  $_{-0.04}^{+0.03}$  & -0.31 & . . . & . . . &  m28 / a,c &  17 \\
SL\,716  & 0.20 & 2.6 & 0.068 &  0.28  $_{-0.03}^{+0.04}$  & -0.10 & . . . & . . . &  m28 / a,c &  17 \\
BSDL\,3001  & 0.77 & 2.5 & 0.068 &  0.28  $_{-0.03}^{+0.04}$  & -0.31 & . . . & . . . & a,c &  17 \\
BSDL\,2995 & 0.50 & 2.4 & 0.070 &  1.00 $\pm$ 0.20  & -0.40 & . . . & . . . &  m29  &  9 \\
BSDL\,3000  & 0.63 & 2.5 & 0.067 &  0.22  $_{-0.03}^{+0.02}$  & 0.00 & . . . & . . . & c &  17 \\
BSDL\,3003  & 0.41 & 2.6 & 0.068 &  0.45 $\pm$ 0.05  & -0.31 & . . . & . . . & c & 17 \\
H88-334 & . . . & 2.5 & 0.072 & . . . & . . . & 2.00 & . . . & . . . & 1 \\
BSDL\,3050  & 0.34 & 2.5 & 0.067 &  0.28  $_{-0.03}^{+0.04}$  & -0.31 & . . . & . . . &  m30  &  17 \\
KMHK\,1389  & 0.36 & 2.6 & 0.067 &  0.16 $\pm$ 0.02  & -0.31 & . . . & . . . &  m30  &  17 \\
BSDL\,3060  & 0.50 & 2.6 & 0.067 &  0.45  $_{-0.1}^{+0.11}$  & -0.47 & . . . & . . . & . . . & 7 \\
HS\,420  & 0.41 & 2.6 & 0.073 &  0.32  $_{-0.07}^{+0.08}$  & -0.27 & . . . & . . . & . . . & 7 \\
BSDL\,3072  & 0.41 & 2.6 & 0.073 &  0.40  $_{-0.08}^{+0.16}$  & -0.31 & . . . & . . . &  m30  & 7 \\
BSDL\,3071  & 0.27 & 2.5 & 0.075 &  0.14 $\pm$ 0.20  & -0.41 & . . . & . . . & . . . &  17 \\
KMHK\,1408  & 0.59 & 2.6 & 0.073 &  0.50  $_{-0.1}^{+0.13}$  & -0.33 & . . . & . . . &  m30  & 7 \\
SL\,736  & 0.90 & 2.7 & 0.070 &  0.40  $_{-0.08}^{+0.10}$  & -0.19 & . . . & . . . & . . . & 7 \\
HS\,424  & 0.50 & 2.7 & 0.074 &  0.40  $_{-0.08}^{+0.1}$  & -0.23 & . . . & . . . & . . . & 7 \\
H7   & . . . & 3.2 & 0.065 & . . . & . . . & 1.40 & . . . & . . . &  2 \\
SL\,748  & 0.90 & 2.7 & 0.073 & 0.25 $_{-0.04}^{+0.07}$ & -0.36 & . . . & . . . &  m31 / a &  17 \\
HS\,427  & 0.86 & 2.8 & 0.074 &  0.32 $\pm$ 0.03  & -0.36 & . . . & . . . & . . . &  17 \\
KMHK\,1448  & 0.54 & 2.8 & 0.073 &  0.28  $_{-0.03}^{+0.04}$  & -0.31 & . . . & . . . & . . . &  17 \\
BSDL\,3123  & 0.32 & 2.8 & 0.073 &  0.40  $_{-0.08}^{+0.10}$  & -0.31 & . . . & . . . & . . . & 7 \\
C11  & 0.68 & 3.4 & 0.101 &  0.40 $\pm$ 0.05  & -0.31 & . . . & . . . & . . . &  5 \\
BSDL\,3158  & 0.99 & 3.5 & 0.101 &  2.50 $\pm$ 0.30  & -0.40 & 2.10 & -0.80 & . . . & 5,17 \\
KMHK\,1504  & 0.63 & 3.6 & 0.117 &  2.20  $_{-0.40}^{+0.30}$  & -0.72 & . . . & -0.80 & . . . & 7 \\
SL\,769 & . . . & . . . & 0.076 & . . . & . . . & 1.80 & -0.50 & . . . &  3 \\
H88-365 & 0.41 & 3.5 & 0.101 &  0.28 $\pm$ 0.06  & -0.40 & . . . & . . . & a & 13,17 \\
SL\,775  & 0.95 & 3.5 & 0.090 &  0.63 $_{-0.13}^{+0.17}$  & -0.50 & . . . & . . . & . . . & 7 \\
OHSC\,28   & 0.33 & 8.3 & 0.040 &  2.20 $\pm$ 0.25  & -0.70 & 2.70 & . . . & . . . & 6 \\
NGC\,2161  & 1.50 & 5.9 & 0.130 &  1.10 $\pm$ 0.30  & -0.70 & . . . & . . . & . . . &  14 \\
NGC\,2153  & . . . & 4.6 & 0.035 & . . . & . . . & 1.30 & . . . & a &  2 \\
NGC\,2155  & . . . & 5.4 & 0.050 &  3.20 $\pm$ 0.60  & -0.90 & 3.60 & . . . & . . . &  15 \\
SL\,817 & . . . & 3.6 & 0.070 & . . . & . . . & 2.5 / 1.5 & -0.50 & . . . &  2,3 \\
SL\,826  & 1.17 & 4.4 & 0.112 &  2.50 $_{-0.30}^{+0.70}$ . & -0.78  & 2.10 & -0.90 & . . . &  17 \\
ESO\,121-03   & . . . & 10.4 & 0.030 & . . . & . . . & 8.50 & -1.05 & . . . &  2,3 \\
LW\,393 & . . . & 4.6 & 0.116 & . . . & . . . & 1.80 & . . . & . . . & 1 \\
LW\,397 & . . . & 4.9 & 0.109 & . . . & . . . & 1.80 & . . . & . . . & 1 \\
SL\,842 & 0.40 & 8.1 & 0.030 & . . . & . . . & 1.90 / 2.20  & -0.60 & . . . &  2,3 \\
KMHK\,1668 & . . . & 4.9 & 0.109 & . . . & . . . & 1.70 & . . . & . . . &  1 \\
NGC\,2213  & . . . & 4.6 & 0.116 & 1.5 & -0.40 & . . . & . . . & . . . &  16 \\
SL\,862 & . . . & 4.7 & 0.090 & . . . & . . . & 1.80 & -0.85 & . . . &  2,3 \\
SL\,870  & 1.04 & 5.4 & 0.088 &  1.25  $_{-0.1}^{+0.2}$  & -0.40 & 1.20 & . . . & . . . & 5 ,17\\
SL\,869 & . . . & 4.9 & 0.101 & . . . & . . . & 1.70 & . . . & . . . & 1 \\
KMHK\,1702 & 0.45 & 5.3 & 0.110 &  1.12 $\pm$ 0.10  & -0.72 & 1.20 & . . . & . . . & 5,17 \\
OHSC\,33 & . . . & 6.2 & 0.090 & . . . & . . . & 1.20 / 1.40 & -1.00 &  m32 / c & 2,3 \\
SL\,874 & 0.84 & 4.9 & 0.090 &  1.50 $\pm$ 0.30  & -0.70 & . . . & . . . & . . . &  14 \\
KMHK\,1719 & 0.42 & 5.1 & 0.090 &  1.40 $\pm$ 0.30  & -0.60 & . . . & . . . & . . . &  14 \\
LW\,469 & 0.50 & 5.9 & 0.080 &  0.60 $\pm$ 0.10  & -0.40 & . . . & . . . & . . . &  9 \\
SL\,896 & . . . & 6.4 & 0.070 &  2.30 $\pm$ 0.30  & -0.60 & 2.30 & . . . & . . . &  15 \\
OHSC\,37 & . . . & 9.4 & 0.150 & . . . & . . . & 2.70 / 2.10 & -0.65 & . . . & 2,3 \\
\hline
%\addtocounter{table}{1}
\end{longtable}
\tablefoot{Parameters obtained also by (a) \citet{glatt}, (b) \citet{popescu}, and (c) \citet{pu00}\\
The letter m indicates that the cluster belongs to a binary or multple  system.}
\tablebib{ 
1) \citet{p11}; 2) \citet{g97}; 3) \citet{b98}; 4) \citet{choud}; 5) \citet{palma13} ; 6) \citet{p09b}; 7) \citet{palma15}; 8) \citet{p14}; 9) \citet{p12a}; 10) \citet{p03a}; 11)  \citet{p03b}; 12) \citet{g03}; 13) \citet{palma11}; 14) \citet{p11b}; 15) \citet{p02}; 16) \citet{g87}; 17) Present work  } 
%\end{longtab}

\twocolumn

\end{document}